\PassOptionsToPackage{table}{xcolor}

\documentclass[sigconf,authorversion,nonacm]{acmart}

\AtBeginDocument{%
  \providecommand\BibTeX{{%
    \normalfont B\kern-0.5em{\scshape i\kern-0.25em b}\kern-0.8em\TeX}}}

\usepackage[capitalise]{cleveref}
\usepackage{algorithm}
\usepackage{algpseudocode}
\usepackage{pgfplots}
\usepackage{tikz}

\usepackage{subcaption}
\pgfplotsset{width=5.5cm,compat=1.9}
\usepackage{multirow}
\usepackage{makecell}
\usepackage{color, colortbl}
\usepackage[flushleft]{threeparttable}
\usepackage{adjustbox}
\usepackage{float}
\definecolor{Gray}{gray}{0.9}
\definecolor{Blue}{rgb}{0.84,0.92,0.94}
\usepackage{tabularx}
\usepackage{changepage}
\usepackage[table]{xcolor}
\DeclareUnicodeCharacter{2212}{-}
\usepackage{booktabs}

\definecolor{mycolor}{rgb}{0.122, 0.435, 0.698}
\makeatletter
\newcommand{\mybox}[1]{%
  \setbox0=\hbox{#1}%
  \setlength{\@tempdima}{\dimexpr\wd0+13pt}%
  \begin{tcolorbox}[colframe=mycolor,boxrule=0.5pt,arc=4pt,
      left=6pt,right=6pt,top=6pt,bottom=6pt,boxsep=0pt,width=\linewidth-4pt,center]
    #1
  \end{tcolorbox}
}
\makeatother
\begin{document}

\title{Redundancy and Concept Analysis for Code-trained Language Models}

\author{Arushi Sharma}
\affiliation{%
  \institution{Iowa State University}
  \country{USA}}
\email{arushi17@iastate.edu}

\author{Zefu Hu}
\affiliation{%
  \institution{Iowa State University}
  \country{USA}}
  \email{zefuh@iastate.edu}
  
\author{Christopher J. Quinn}
\affiliation{%
  \institution{Iowa State University}
  \country{USA}}
  \email{cjquinn@iastate.edu}

\author{Ali Jannesari}
\affiliation{%
  \institution{Iowa State University}
  \country{USA}}
  \email{jannesar@iastate.edu}

\renewcommand{\shortauthors}{Sharma, et al.}

\begin{abstract}
Code-trained language models have proven to be highly effective for various code intelligence tasks. However, they can be challenging to train and deploy for many software engineering applications due to computational bottlenecks and memory constraints. Implementing effective strategies to address these issues requires a better understanding of these 'black box' models. In this paper, we perform the first neuron-level analysis for source code models to identify \textit{important} neurons within latent representations. We achieve this by eliminating neurons that are highly similar or irrelevant to the given task. This approach helps us understand which neurons and layers can be eliminated (redundancy analysis) and where important code properties are located within the network (concept analysis). Using redundancy analysis, we make observations relevant to knowledge transfer and model optimization applications. We find that over 95\% of the neurons are redundant with respect to our code intelligence tasks and can be eliminated without significant loss in accuracy. We also discover several subsets of neurons that can make predictions with baseline accuracy. Through concept analysis, we explore the traceability and distribution of human-recognizable concepts within latent code representations which could be used to influence model predictions. We trace individual and subsets of important neurons to specific code properties and identify  'number' neurons, 'string' neurons, and higher-level 'text' neurons for token-level tasks and higher-level concepts important for sentence-level downstream tasks. This also helps us understand how decomposable and transferable task-related features are and can help devise better techniques for transfer learning, model compression, and the decomposition of deep neural networks into modules.

\end{abstract}




\maketitle

\section{Introduction}

Code-trained language models are driving the rapid advancement of data-driven software engineering. Their effectiveness has significantly reduced the cost of feature engineering. Software engineering (SE) tools for various code intelligence tasks like defect detection, clone detection, code summarization, code generation, etc. are increasingly powered by large neural networks (eg. OpenAI CodeX for Github Copilot). Neural networks, being universal function approximators \cite{hornik1989multilayer} with millions or billions of parameters, have a high capacity to generalize to different downstream tasks. However, their opacity makes it hard to trust their predictions or devise effective strategies to mitigate challenges due to their size. Moreover, large neural models often require huge datasets and processing power. Manually annotating datasets may not be feasible most of the time, and generating synthetic datasets, though efficient, can lead the deep learning model to memorize spurious correlations \citep{rabin2023memorization}. In the context of software engineering tasks like vulnerability detection, a neural model relying on spurious correlations to make predictions about the presence of bugs in code can have major security implications.


Additionally, the rapid deployment of code-trained language models necessitates condensing these models to reduce inference time computation cost for deployment on memory-constrained edge devices and to achieve reduced latency over distributed architectures. Research into techniques like knowledge transfer and model compression to address such computational bottlenecks is still in the early stages for source code models \citep{feng2020codebert,guo2022unixcoder,lu2021codexglue, chen2022neural,pal2022cross,jiang2023knod} and can benefit from a better understanding of these models. Although a few studies have aimed to improve the interpretability of code-trained language models, exploring concerns like vulnerability signal traceability \citep{rabin2021understanding,rabin2022syntax}, explainable automated program repair \citep{kang2023explainable}, and comparing different source code embeddings \citep{9609166,kang2019assessing,siow2022learning,rabin2020towards,karmakar2021pre,troshin2022probing,wan2022they,lopez2022ast} for various tasks, we are still not close to developing reliable and explainable code intelligence models and interpretability studies that can guide knowledge transfer and model optimization decisions.


In this paper, we introduce a novel approach to improving the interpretability of code-trained language models by conducting a fine-grained neuron-level interpretability study on the latent representations learned by these models. Our approach is adapted for source code models from neuron analysis approaches in Natural Language Processing \citep{dalvi2020analyzing,durrani2022linguistic,durrani2020analyzing,dalvi2019one}, as discussed in Section \ref{Approach}.

Our paper makes the following contributions:
\begin{enumerate}
\item We perform the first neuron-level analysis on seven different code-trained language models with respect to four different software engineering tasks. We use different neuron ranking and selection techniques to remove redundancies and identify important neurons with respect to a given task.

\item We use neuron analysis techniques discussed in Section \ref{Approach-Neuron Analysis Techniques} to eliminate neurons that are either very similar to each other or are irrelevant with respect to a given task. We discuss implications in knowledge transfer, model optimization, and decomposability of neural networks.

\item We systematically study the concentration and distribution of the \textit{important} neurons with respect to concepts or code properties learned by the latent representations of code-trained models. We discuss implications in traceability between inputs, neurons, and model outputs.
\end{enumerate}

This paper is organized as follows: Section \ref{Related Work} discusses related work, Section \ref{Approach} describes our approach and the different techniques used to perform our neuron analysis, followed by experimental settings and evaluation in Section \ref{Evaluation}. \cref{Threats to Validity} explores the threats to validity, and Section \ref{Conclusion and Future work} concludes this paper and provides directions for future work.

\begin{figure*}[h]
    \caption{Workflow diagram} \label{fig: Workflow}
    \centering
	\includegraphics[width=13cm]{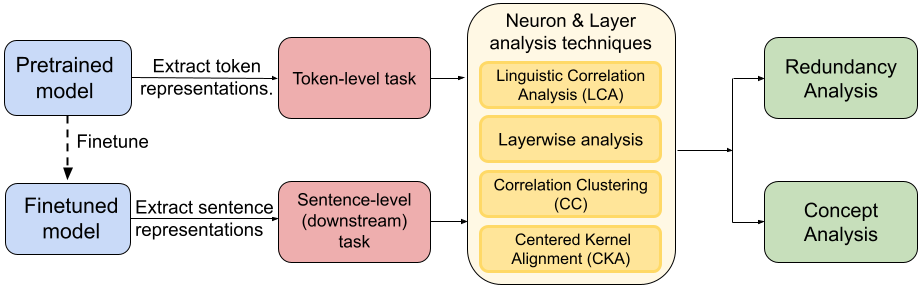}
\end{figure*}


\section{Related Work}
\label{Related Work}

In Sections \ref{Related Work-Extractive} and \ref{Related Work-embeddings}, we discuss the two major directions that post-hoc interpretability research for code-trained models has advanced in. Section \ref{Related Work - Knowldege transfer and model optimization} discusses knowledge transfer and model optimization research for neural code intelligence models. 

\subsection{Extracting Input Features to explain model predictions}
\label{Related Work-Extractive}

The first direction has used \textit{extractive} interpretation approaches which extract and highlight key regions of the input program and allow the user to draw conclusions about corresponding model predictions. Their rationale is that studying the impact of input code reduction on model predictions can improve \textit{traceability} between relevant code features and model outputs and provide better insights into the black box models. Studies in extractive interpretability have used techniques like attention-guided code perturbation \citep{bui2019autofocus, shen2022benchmarking,zhang2022diet}, prediction preserving code minimization using syntax unaware  approaches \citep{suneja2021probing, rabin2021understanding}, and syntax aware approaches \citep{rabin2022syntax}.   
Attention-based code perturbation techniques are useful when extracting hard-coded features is not a viable option for dynamically typed languages or arbitrary code-snippets. However, not only do most attention-based studies poorly correlate with code elements that are important according to human programmers \citep{rabin2022syntax,paltenghi2021thinking}, the debate regarding whether attention is an explanation  \citep{jain2019attention,wiegreffe2019attention} and whether it aligns with information stored within learned embeddings \citep{wan2022they}, whether that information can influence model predictions  is still open \citep{bui2019autofocus}. Therefore, attention is best suited to guiding interpretability studies in combination with other techniques to provide a holistic picture of the inner workings of neural models \citep{wiegreffe2019attention}. For example, \citet{wan2022they} uses attention analysis to see whether the model focuses on the syntax structure of code, followed by probing analysis to ascertain whether the information was encoded within the latent representations. 
 \citet{bui2019autofocus} uses attention-guided code perturbation to rank input elements based on their influence on model predictions. On the other hand, prediction preserving input minimization approaches in software engineering have been used to extract meaningful input features \cite{rabin2021understanding,rabin2022featureextractor,rabin2022syntax,suneja2021probing,cito2022counterfactual}. These approaches work particularly well when features are sparse and uncorrelated so that highlighting specific parts of the input can give clear indications as to it role in the model prediction \citep{rabin2021understanding}. They rely on perturbing code elements using syntax-unaware techniques like delta-debugging \cite{ rabin2021understanding,suneja2021probing} while  \citet{rabin2022syntax} provides a syntax-aware approach for program simplification to improve understanding about model predictions. Program simplification or removal of tokens while retaining output predictions can lead to changes in attention weights and activations specially because many BERT and GPT2-based models use positional encodings. \citet{cito2022counterfactual} take this further by generating counterfactual explanations for model predictions. 
 
 These methods are model-agnostic and can be adopted without knowledge of model internals, which may not be accessible if the models are not open-sourced. On the other hand, without information about models latent representations, they provide limited insight into the general effectiveness of code embeddings and the distribution of information encoded within them. Additionally,  \citet{rabin2021understanding} suggested  that ``models do not necessarily rely on overall structure or content of code but on very few features leveraging simple syntactic patterns to make predictions.'' Source code models tend to rely on very few key input tokens \citep{rabin2021understanding}, mainly focus on keywords and datatypes, and over rely on syntactic features like semicolons \citep{zhang2022diet} often at the cost of indicators that are more important according to human programmers. That source code models are prone to learning spurious correlations between superficial input features \citep{yefet2020adversarial} makes them susceptible to adversarial examples that can influence model predictions and cause loss of generalizability in the face of even small semantic transformations \citep{rabin2021generalizability}. This motivates our research into identifying individual and subsets of neurons responsible for specific code properties and improving traceability between neurons and input features.


\subsection{Interpreting source code embeddings}
\label{Related Work-embeddings}

The second direction of interpretability research for source code models has focused on investigating models' latent representations and specific properties of code encoded within them. The goal of these studies is to explain general model behavior with the goal of increasing transparency of these `black box' neural models to guide model pretraining and optimization. To this end, \citet{9609166} assess the generalizability of CodeBERT on various software engineering tasks while \citet{kang2019assessing} assess the effectiveness of the Code2Vec embedding in different tasks. \citet{siow2022learning} performs an empirical study to evaluate different types of code embeddings over several machine learning models. 
\citet{rabin2020towards} look at the internals of Code2Vec source code embedding vectors by using them to train SVM models and comparing them with SVM models trained on handcrafted features. 
There has also been research using probes to ascertain whether pretrained models capture  syntactic, structural, and semantic information about source code \citep{karmakar2021pre, troshin2022probing, wan2022they, lopez2022ast}. \citet{troshin2022probing} has compared several models on the basis of architecture, finetuning, model size, and code-specific pretraining objectives. The performance of probing classifiers on diagnostic tasks to predict AST depth, whether a token is an identifier, data flow edges, variable names, etc., have indicated that the code embeddings obtained using neural models of code do capture syntactic and semantic code properties to some extent. Further, \citet{karmakar2021pre} uses a `naive' baseline which refers to the performance of the neural model's embedding layer on the probing task, and \citet{troshin2022probing} have used a ``simple bound baseline'' with which to evaluate their probing classifiers. However, these do not account for accuracies that may be inflated due to probe memorization. We introduce the role of memorization in probing classifiers due to the nature of source code datasets to source code model probing and propose Selectivity-guided probing task formulation to mitigate its effects. 

Additionally, \citet{lopez2022ast} introduces an AST probe that can recover the entire Abstract Syntax Tree of an input code snippet from the latent representations of code-trained language models. \citet{wan2022they} use attention analysis to analyze whether syntactic relationships are captured by the model, followed by using structural probes to see whether that information was actually captured within the learned embeddings. They further induce a syntax tree to conclude that the syntactic structure of code is captured by the pretrained model. While related work has only used probing classifiers to claim whether or not a certain property is encoded within the latent representations, our work uses the probing classifier-based neuron ranking method, Linguistic Correlation Analysis to eliminate redundant neurons and isolate important neurons wrt a given task. Our neuron analysis attempts to improve traceability between input features and features encoded within latent representations and study its role in impacting model behavior.

\subsection{Knowledge transfer and model optimization for code-trained language models}
\label{Related Work - Knowldege transfer and model optimization}
 Code-trained language models often require large datasets. This can be particularly hard to come by for specialized software engineering tasks that may need manual annotation. To avoid this overhead, software engineers often generate synthetic datasets. However, the process of generating synthetic datasets may induce spurious correlations that the deep learning model learns. Therefore, \citet{chen2022neural} propose to use transfer learning to address the challenge of training a deep learning model for one task (vulnerability fixing), for which only a small dataset is available---first training the network on a related task (bug fixing) for which a large data set was available and then fine-tuning the network on the primary task (vulnerability fixing). \citet{pal2022cross} perform a systematic literature review of cross-project defect prediction approaches which use historical data collected from previous versions of one project to predict faulty modules in another project. \citep{sharma2021code} use transfer learning for detecting code smells which are problems in source code that will compile as intended but do not follow correct design principles and may cause issues further down the line.  Further, \citep{pan2020decomposing}  
 propose decomposing monolithic deep learning networks into reusable modules to facilitate transfer learning as well as reducing memory and computational requirements through model compression. Additionally, \citep{jiang2023knod} performs domain-knowledge distillation (KNOD) for automated program repair to enforce syntactic and semantic rules during training and inference. They use a teacher network trained over syntactic and semantic rules to teach a student network (the encoder-decoder model for the automated program repair task to take code structure into account. Insights obtained from our study about the transferability and decomposability of encoded knowledge can be used to guide such  knowledge transfer and model compression approaches.

\section{Approach}
\label{Approach}


\subsection{Definition of concepts}

\subsubsection{Redundancy} 
\label{Approach-Definition-Redundancy}
\citet{dalvi2020analyzing} shows that pretrained language models in Natural Language Processing (NLP) contain large amounts of redundant information. These redundancies may arise due to two reasons. First,  the models are extremely large and tend to be overparameterized, leading to overlapping information in different parts of the network.  Second, the models are trained on general objectives and may contain information that is redundant with respect to a particular downstream task. We eliminate redundancies based on similarity and task relevance in code-trained models and isolate specific subsets of neurons that can perform well on the given task, allowing us to draw conclusions about the distribution of information in these deep networks.

\subsubsection{Concept learning}
\label{Approach-Definition-Concept Learning}
Concept learning is used to analyze the ability of deep neural networks to encode different linguistic and non-linguistic concepts like parts of speech, morphology, etc, to interpret their inner mechanics. In this paper, we define a concept as a human-recognizable property of the input code encoded within latent representations of the models.  We focus on pre-defined concepts wrt. the objective of our tasks. In the case of the Token-Tagging task, the concepts are classes like IDENTIFIER, KEYWORD, STRING, etc., and we study how these concepts are encoded within the models. In case of our downstream tasks which are sentence-level tasks, we analyze the distribution of more condensed higher-level concepts on models finetuned for these tasks. 

When analyzing the distribution of important neurons, we refer to the following terms as defined by \citep{gurnee2023finding}. 
Neurons that contain a 1:1 mapping between features of the input and neurons in a network are referred to as \textit{monosemantic Neurons}. This is most obvious when the number of neurons equals the number of neurons. \textit{Polysemantic neurons} are those which can activate for a large number of seemingly unrelated properties and are a result of superposition. \textit{Superposition} occurs when the number of features is larger than the number of neurons, and the model compresses multiple features into smaller dimensions for greater representational power. \textit{Composition} occurs when several neurons are responsible for a single property and is our focus in this study. We do not discuss these concepts in detail because confirmation is either methodologically unfeasible or requires further experiments. We list our observations to motivate further exploratory research.




\subsection{Code-trained language models}

\begin{table}[!ht]
  \centering
  \caption{Models}
  \label{tab:Models}
  \small 
  \begin{tabular}{l|c|c|c|c}
    \toprule
    Model & Arch & Pretraining Obj & Modality & \makecell{Pretraining\\ Data } \\
    \midrule
    BERT\cite{devlin2018bert} & Enc & MLM, NSP & NL & English \\
    \hline
    \makecell{RoBERTa\\(RBa)\cite{liu2019roberta}} & Enc & MLM, NSP & NL & English \\
    \hline
    \makecell{CodeBERT \\
    (CB)\citep{feng2020codebert} }& Enc & MLM, RTD & \makecell{PL \& \\ (NL + PL)} & Multi \\
    \hline
    GCB\cite{guo2020graphcodebert} & Enc & MLM, EP, NA & \makecell{NL + PL \\ + DFG (CS)} & Multi \\
    \hline
    \makecell{UniXCoder\\ (UC)\cite{guo2022unixcoder} }& Enc & \makecell {MLM, MCL, \\ CMG} & \makecell{ NL comments \\ + CS (AST)} & Multi \\
    \hline
    CGP\cite{lu2021codexglue} & Dec & ULM & PL & Python \\
    \hline
    CGJ\cite{lu2021codexglue} & Dec & ULM & PL & Java \\
    \bottomrule
  \end{tabular}
  
  \begin{tablenotes}
    \small
    \item \textbf{Architecture} Enc: Encoder, Dec: Decoder \textbf{Pretraining Objectives} MLM: Masked Language Modeling, NSP: Next Sentence Prediction, RTD: Replaced Token Detection, EP: Edge Prediction, NA: Node Alignment, MCL: Multi-modal Contrastive Learning, CMG: Cross Modal Generation, ULM: Unidirectional Language Modeling. Details and loss functions for each objective can be found in the supplementary material. \textbf{Modality} NL: Natural Language, PL: Programming Language, CS: Code Structure. 
  \end{tablenotes}
\end{table}


We conduct experiments on seven code-trained language models (Table \ref{tab:Models}). These models have been widely used in various software engineering tasks and were selected based on differences in architecture, pretraining objectives, data, modalities, etc. \citep{niu2023empirical}, providing diverse points of comparison for our neuron-level analysis.

Although large language models like GPT-4, LLaMA, BARD, etc., are gaining popularity, performing interpretability studies on our selected models is still valuable for the following reasons: (i) Conducting white-box interpretability studies by extracting activations of several GPT-4-sized models is not feasible. Insights gained from our studies can be useful for knowledge transfer or guiding black-box interpretability studies on larger models. (ii) Having fewer points of comparison allows us to make more concrete observations. (iii) Encoder-based models perform well for some classification tasks. \citep{niu2023empirical} demonstrated that certain encoder models, such as UniXCoder, remain state-of-the-art for specialized classification tasks. (iv) There is consistency in the architectures of generative models. While scaling may introduce new challenges due to increased complexity and the number of parameters, studying GPT2 models can provide foundational understanding and insights into how interpretability scales with model size.

\subsection{Probing neural code embeddings}

\label{Background-Probing Classifiers}
We use probing classifiers are a well-established technique to probe latent representations in NLP neural models and have also seen several applications in source code model interpretability. 
 
Let  $\mathbb{M}$ 
denote a code-trained language model, a function that maps an input $x$ (i.e. a line of source code or entire code snippet) to a vector $\mathbb{M}(x)$ of neuron activation values (latent representations). A probing classifier $g : \mathbb{M}(x) \mapsto \hat{t}$ maps those latent representations $\mathbb{M}(x)$ to some property $\hat{t}$ of interest  (e.g., token label).
The probing classifier $g$ is trained by minimizing the elastic-net regularized categorical cross-entropy loss \citep{zou2005regularization}. The regularization hyper-parameters can be selected through a grid search using cross-validation, though the computation required for such searches can be prohibitive, so it is not uncommon to use manually selected values.     

\subsection{Selectivity-guided probing-task formulation}
\label{Approach-Selectivity guided probing task formulation}

Nominally, one could use the test performance of a probing classifier on a source code property task to assess the extent to which that property is learned during pre-training by BERT or GPT-based source code models. Previous work in probing source code models \citep{karmakar2021pre,troshin2022probing} has used the accuracy metric to evaluate probe performance on a given probing task. However, in the NLP literature, probing classifiers were in some cases found to achieve high test accuracy but essentially did so through memorization.  To mitigate against the potential for memorization, \citet{hewitt2019designing} introduced control tasks that randomly assign a new label for each token, preserving label frequencies. They argue that a high control task accuracy highlights the role of training data and the probe's memorization capabilities instead of using encoded information about linguistic structure and semantics. We use this methodology to create a control task for our Token Tagging task. Our control task is defined by mapping each token type $t_i$ to a randomly selected class to a randomly sampled behavior $C(t_i)$, from a set of numbers $\{1 \dots T\}$ where $T$ is the number of classes or labels. \citet{hewitt2019designing} introduced the \textit{Selectivity} metric, the difference between linguistic task accuracy and control task accuracy. We use this metric to put probing task accuracy in context with the probe’s capacity to memorize from token types and detect information leakage between source code training and test sets. An effective probing task is one with high linguistic task accuracy and low control task accuracy. We use linear classifiers for our probing experiments which are known to be the most selective \cite{hewitt2019designing}. 

Further, \citet{casalnuovo2019studying,rabin2021generalizability} find that model outputs are frequently affected by the repetitive nature of source code and code duplication due to code mined from open source Github repositories\cite{allamanis2019adverse}. Our experiments (section \ref{Evaluation-Experimental Settings}) suggested that a large amount of memorization in source code data is due to the repetitive and deterministic nature of the tokens. This is particularly important for token-level tasks as sentence-level tasks are probed using sentence embeddings that have been optimized by finetuning. They presumably store higher-level information about the downstream task. Therefore, We reduce the overlap between the training and the test set of our token-tagging task by eliminating all deterministic tokens like EQUAL "=", LPAR ")", etc., and probe the model on just six classes with ambiguous tokens IDENTIFIER, KEYWORD, STRING, MODIFIER, TYPE, and NUMBER. IDENTIFIER includes all method names, variable names, constants, etc. KEYWORD includes all Java keywords, for example, "False", "await", and so on; STRING includes all strings in Java source code; these are usually found within quotes; MODIFIER includes "static", "protected", "private", and so on; TYPE includes "boolean", "float", "int" and so on; and NUMBER which includes all numbers found within a Python code snippet.

\begin{figure*}[h]
    \caption{Control Tasks and Selectivity baseline} \label{fig: Selectivity}
    \centering
	\includegraphics[width=14cm]{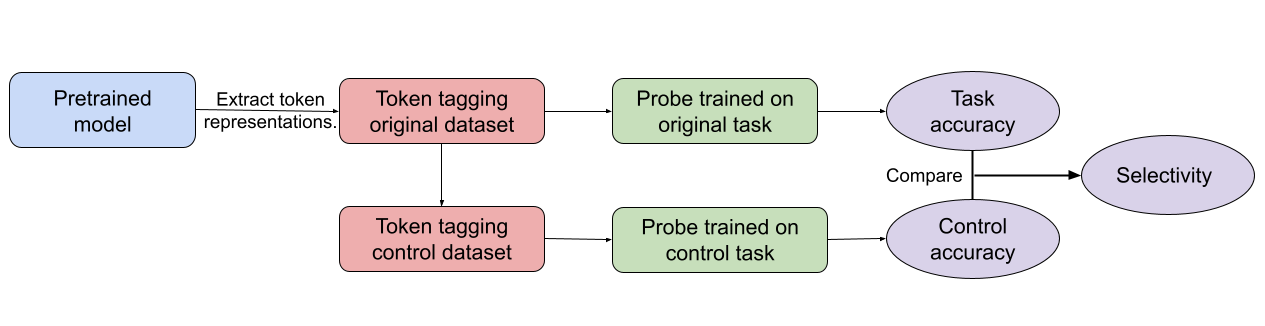}
\end{figure*}



\subsection{Neuron Analysis Techniques}
\label{Approach-Neuron Analysis Techniques}
We perform a comprehensive analysis using different neuron analysis techniques to identify important neurons. \cref{fig: Workflow} shows our workflow for a neuron-level analysis. 


\subsubsection{Probe-Based Ranking of Neurons Using Linguistic Correlation Analysis}
\label{Background-Linguistic Correlation Analysis}
We use Linguistic Correlation Analysis (LCA) \cite {dalvi2019one,durrani2022linguistic} to extract important neurons with respect to a given task. We train 
probing classifier $g$  on the neuron activations of model $\mathbb{M}$. We then use the absolute values of the weights, $\theta\in\mathbb{R}^{N\times|T|}$ of the probing classifier $g$, as a measure of the importance of neurons and select the top $k$ neurons with respect to a given property $t \in T$. The algorithm is described in the supplementary material.


\subsubsection{Correlation Clustering}
\label{Background-Correlation Clustering}

Correlation Clustering is task-independent and can be used to eliminate neurons that are redundant because they capture highly similar information. We perform Correlation Clustering \citep{bansal2004correlation} of neurons based on their activation patterns. We calculate Pearson's correlation coefficient for every neuron pair over their activation values for the samples in the data set $D$. Next, with the distance matrix $cdist(x, y) = 1 - |corr(x, y)| $ we use  agglomerative hierarchical clustering  with average linkage to minimize the average distance of all data points in pairs of clusters. We  randomly selected one neuron from each cluster to create a reduced set of \textit{independent neurons} which are then used to train a probing classifier.
The clustering threshold hyperparameter $c$ can range from 0 to 1 and controls the maximum distance between any two points in a cluster. A high value of $c$  indicates that the clusters are larger and their neurons that are further apart are included in the cluster. 

\subsubsection{Layer Selection}
\label{Background-Layer Selection}
We perform independent and incremental layerwise analysis by selecting activations of neurons from specific layers. Independent layerwise analysis selects neurons from a specific layer $l$  to train a probing classifier. The incremental layer-wise analysis is performed by concatenating the layer probing classifiers for every layer before the given layer i.e. layer 0 to layer $l$. Layerwise analysis can be used to identify redundant layers with respect to the given task.  

\subsubsection{Minimal Neuron Set}
\label{Background-Minimal neuron set}
The minimal neuron set is the smallest number of neurons that can be obtained using a combination of our neuron selection methods and can maintain the accuracy of the probing classifier $g$ using all neurons, which we refer to as the  \textit{oracle accuracy}. We use Layer Selection, followed by Correlation Clustering and Linguistic Correlation Analysis to select the minimal neuron set. The idea is to select the set of lowest layers that give the best accuracy using incremental layerwise results, to eliminate highly similar neurons using correlation clustering, and then rank the remaining neurons using LCA to obtain a minimal neuron set.  

\subsubsection{Centered Kernel Alignment (CKA)}
Centered Kernel Alignment (CKA) \cite{kornblith2019similarity} is a similarity metric based on the Hilbert-Schmidt Independence Criterion (HSIC) \cite{gretton2007kernel} and is used to measure the similarity between representations of features in neural networks. The method is popular due to its invariance  to invertible orthogonal transformation and isotropic scaling.  We use it to compare adjacent layers within our networks in order to draw conclusions about the architectural choices of the network.  For token-tagging, we randomly select 25,000 tokens out of 580,000 tokens to calculate the CKA values. For all sentence-level tasks, we randomly select 25,000 sentences.

\section{Evaluation}
\label{Evaluation}

\subsection{Experimental Settings}
\label{Evaluation-Experimental Settings}

\paragraph{Models and Tasks}
\label{Experiments-Models}
The models we used can be found in table \ref{tab:Models}. We used pretrained models from the huggingface transformers library \citep{wolf-etal-2020-transformers}. We perform our experiments on four tasks. Data statistics for our tasks can be found in table \ref{tab:dataset_statistics}. The details of different tasks, and dataset preprocessing are provided in the supplementary material. Our source-code is available on github\footnote{\url{https://anonymous.4open.science/r/interpretability-of-source-code-transformers-0F51/README.md}}.

\paragraph{Finetuning and extracting neuron activations}
For the token-level task of Token-Tagging, we extract the neuron activations for each token in our data set.  For sentence-level tasks, we first finetune the models using the hyperparameters provided by \cite{lu2021codexglue} for finetuning our models, update model parameters using the Adam optimizer and then extract sentence level representations. The fine-tuning step is essential to optimize the first token <s> as it is the aggregate pooling of the entire sentence in BERT-like models. For code search which is formulated as sentence-pair classification tasks, we extract the <s> token for each snippet.  For GPT2-based models, which are trained on the unidirectional language modeling task, we extract the last token. Since GPT2 models pad to the right, we extract the end-of-sentence token </s> to obtain sentence-level representations.  

\paragraph{Probing classifier settings}

The activations of the original training set for a given task are split into a new training and validation set in a 90:10 ratio, while the original validation set is used as the test set for the the probing experiments. We use the NeuroX\footnote{\url{https://github.com/fdalvi/NeuroX}} \citep{dalvi2019neurox} toolkit and adapt \citet{dalvi2020analyzing}'s experiments\footnote{\url{https://github.com/fdalvi/analyzing-redundancy-in-pretrained-transformer-models}} to perform the probing and neuron- ranking experiments for code-trained models. Our probe is a logistic regression classifier with ElasticNet \cite{zou2005regularization} regularization using a categorical cross-entropy loss 
optimized by Adam \citep{kingma2014adam}.  We train the probing classifiers for 10 epochs with a learning rate of 1e−3, batch size of 128 and regularization hyper-parameter  $\lambda_1=\lambda_2=10^{-5}$ for both the $L_1$ and $L_2$ regularization penalties. 
We used the default hyperparameter values suggested in \citep{dalvi2020analyzing}
\footnote{\url{https://github.com/fdalvi/analyzing-redundancy-in-pretrained-transformer-models}}. 

\begin{table}
  \caption{Neuron activations dataset statistics}
  \label{tab:dataset_statistics}
  \begin{tabular}{l|l|l|l|l}
    \toprule
    \textbf{Task} &\textbf{Train} & \textbf{Dev} & \textbf{Test}& \textbf{Tags} \\
     \midrule
       \makecell{Token-tagging (filtered)} & 12000 & 1620 & 2040 & 6  \\ 
     \hline
      Defect detection \citep{zhou2019devign} &  19668 & 2186 & 2,732  & 2 \\
      Clone Detection \cite{svajlenko2015evaluating} &  81029 & 9003  & 10766 & 2 \\
       Code Search \cite{Huang2020CoSQA20} & 
	18000 & 2000 & 604 &2 \\
    \bottomrule
  \end{tabular}
\end{table}

\subsubsection{Baselines}
\label{baseline}
We introduce control tasks and the Selectivity\cite{hewitt2019designing} metric to source-code model probing to evaluate the probe's capacity to memorize in token-level tasks and account for artificially inflated accuracies. This is further explained in \cref{Approach-Selectivity guided probing task formulation}.
For our neuron-level analysis, we use a probing classifier trained on the activations of all neurons of a given model as a baseline. We call this \textit{Oracle accuracy}. We further compare our probing classifier-based neuron ranking results with Probeless ranking (see \cref{Background-Probeless Ranking}) to account for the inherent shortcomings of probing classifiers as discussed in \citep{belinkov2022probing}. We also analyze our results against parallel studies in NLP \cite{dalvi2020analyzing} on comparable tasks like Parts of Speech, Sentiment Analysis (SST-2 \cite{socher2013recursive}), Multi-Genre Natural Language
Inference (MNLI \cite{williams2017broad}), Question-Answering Natural Language inference (QNLI \cite{wang2018glue}
[Wang et al. 2018]) due to similarities in downstream task formulation.

\subsubsection{Selectivity-guided java token tagging probing task}

\begin{table}[]
  \caption{Probing results for Java Token Tagging }
  \label{tab:token-probe}
  \begin{tabular}{llcc}
    \toprule
    \textbf{Dataset } & \textbf{ Criterion} & \textbf{CodeBERT} & \textbf{GraphCodeBERT}\\
    \midrule
      \multirow{3}{*} {\textbf{Original}} & Orig. acc. & 0.99 & 0.99\\
      & Control acc. & 0.77 & 0.78 \\
      & Selectivity  & 0.22 & 0.21 \\
      \hline
      \multirow{3}{*} {\textbf{Filtered}} & Orig. acc. & 0.86  & 0.88 \\
      & Control acc. & 0.15 & 0.16 \\
      & Selectivity  & 0.71 & 0.72 \\
  \bottomrule
\end{tabular}
\end{table}


We perform selectivity-guided probing task formulation and filter the dataset as discussed in section \ref{Approach-Selectivity guided probing task formulation}. The dataset contains 6 classes  MODIFIER, INTENTIFIER, KEYWORD, TYPE, NUMBER, and STRING. 
The results of these experiments for two models, CodeBERT and GraphCodeBERT, can be found in \cref{tab:token-probe}.
Our probing experiments reveal a  high control task accuracy of about 77\% for token-tagging,  which suggests that the probe uses memorization to a large extent.  This is because by construction (the random relabelling of the tokens), control tasks discard syntactic code properties (i.e. the patterns of token sequences resulting from the underlying grammar). Control task probes can only perform well (relative to random guessing) if the probe is able to memorize  due to information leakage from the training set to the test set. We addressed this issue of information leakage by creating a filtered version of the Token-Tagging dataset. While using appropriate comparative baselines like \citep{troshin2022probing} may mitigate the impact of inflated accuracies to some extent,  preventing probe memorization due to information leakage is particularly important for our application as we use a probing-based approach to rank important neurons and trace them to specific tokens in the input code. Removing repetitive and deterministic tokens significantly improves selectivity which ensures that probing classifiers are reflective of properties encoded by neural models and do not rely on memorization leading to inaccurate conclusions about model interpretability. The statistics for this filtered data set can be found in \cref{tab:dataset_statistics}. 

\subsection{Redundancy Analysis: Removing similar or task-irrelevant neurons}

\label{Evaluation-Redundancy Analysis}
\begin{table*}[]

\definecolor{lightyellow}{RGB}{255, 255, 204}
\caption{Neuron level analysis} \label{tab: Results1}
\begin{adjustbox}{width=1\textwidth}
\begin{tabular}{|c|c|ccccccc|ccccccc|}
\hline
                               &                        & \multicolumn{7}{c|}{Token Tagging (Java)}                                                                                          & \multicolumn{7}{c|}{Defect Detection}                                                                                              \\ \hline
Selection                      &                        & BERT             & CB               & CGJ              & CGP              & GCB              & RBa              & UC               & BERT             & CB               & CGJ              & CGP              & GCB              & RBa              & UC               \\ \hline
\multirow{2}{*}{Oracle}        & original \# of neurons & 9984             & 9984             & 9984             & 9984             & 9984             & 9984             & 9984             & 9984             & 9984             & 9984             & 9984             & 9984             & 9984             & 9984             \\
                               & Baseline accuracy      & 74.61\%          & 88.63\%          & 85.44\%          & 80.25\%          & 88.38\%          & 85.93\%          & 84.02\%          & 61.46\%          & 64.02\%          & 61.27\%          & 63.58\%          & 63.87\%          & 48.39\%          & 66.80\%          \\ \hline
\multirow{4}{*}{LCA}           & \# of neurons          & 49               & 409              & 369              & 49               & 209              & 229              & 29               & 29               & 79               & 9                & 79               & 489              & 9                & 29               \\
                               & Accuracy               & 79.31\%          & 87.01\%          & 82.25\%          & 80.10\%          & \textbf{88.48\%} & \textbf{87.35\%} & 83.48\%          & 61.57\%          & 64.24\%          & 58.82\%          & 61.60\%          & 62.59\%          & \textbf{56.96\%} & 65.30\%          \\
                               & Diff.                  & 4.70\%           & -1.62\%          & -3.19\%          & -0.15\%          & 0.10\%           & 1.42\%           & -0.54\%          & 0.11\%           & 0.22\%           & -2.45\%          & -1.98\%          & -1.28\%          & \cellcolor{Blue} 8.57\%           & -1.50\%          \\
                               & Neuron reduction       & 99.51\%          & 95.90\%          & 96.30\%          & 99.51\%          & \textbf{97.91\%} & \textbf{97.71\%} & 99.71\%          & 99.71\%          & 99.21\%          & 99.91\%          & 99.21\%          & 95.10\%          & \textbf{99.91\%} & 99.71\%          \\ \hline
\multirow{5}{*}{CC}            & Clustering threshold   & 0.8              & 0.2              & NA               & 0.5              & NA               & NA               & 0.3              & 0.5              & 0.7              & 0.6              & 0.6              & 0.7              & 0.5              & \cellcolor{lightyellow}0.2              \\
                               & \# of neurons          & 116              & 7045             & 9984             & 1102             & 9984             & 9984             & 4430             & 2600             & 1742             & 1023             & 1023             & 2555             & 1977             & 8890             \\
                               & Accuracy               & \textbf{83.82\%} & 87.79\%          & 86.67\%          & 83.09\%          & 88.63\%          & 86.86\%          & 86.03\%          & 62.96\%          & 65.52\%          & 62.59\%          & 62.59\%          & 64.35\%          & 57.50\%          & 67.02\%          \\
                               & Diff.                  & 9.21\%           & -0.84\%          & 1.23\%           & 2.84\%           & 0.25\%           & 0.93\%           & 2.01\%           & 1.50\%           & 1.50\%           & 1.32\%           & -0.99\%          & 0.48\%           & \cellcolor{Blue} 9.11\%           & 0.22\%           \\
                               & Neuron reduction       & \textbf{98.84\%} & 29.44\%          & 0.00\%           & 88.96\%          & 0.00\%           & 0.00\%           & 55.63\%          & 73.96\%          & 82.55\%          & 89.75\%          & 89.75\%          & 74.41\%          & 80.20\%          & 10.96\%          \\ \hline
\multirow{5}{*}{Layerwise(LS)} & Layer Selection        & 0-6              & 0-2              & 0-12             & 0-0              & 0-0              & 0-3              & 0-0              & 0-12             & 0-12             & 0-10             & 0-12             & 0-12             & 0-5              & 0-12             \\
                               & \# of neurons          & 5376             & 2304             & 9984             & 768              & 768              & 3072             & 768              & 9984             & 9984             & 8448             & 9984             & 9984             & 4608             & 9984             \\
                               & Accuracy               & 73.87\%          & 88.58\%          & 82.70\%          & 79.46\%          & 88.53\%          & 85.15\%          & \textbf{94.51\%} & 61.46\%          & 64.02\%          & 62.11\%          & 63.58\%          & 63.87\%          & 57.28\%          & 66.80\%          \\
                               & Diff.                  & -0.74\%          & -0.05\%          & -2.74\%          & -0.79\%          & 0.15\%           & -0.78\%          & 10.49\%          & 0.00\%           & 0.00\%           & 0.84\%           & 0.00\%           & 0.00\%           & \cellcolor{Blue} 8.89\%           & 0.00\%           \\
                               & Neuron reduction       & 46.15\%          & 76.92\%          & 0.00\%           & 92.31\%          & 92.31\%          & 69.23\%          & \textbf{92.31\%} & 0.00\%           & 0.00\%           & 15.38\%          & 0.00\%           & 0.00\%           & 53.85\%          & 0.00\%           \\ \hline
\multirow{6}{*}{LS+CC+LCA}     & Layer Selection        & 0-6              & 0-2              & 0-12             & 0-0              & 0-0              & 0-3              & 0-0              & 0-11             & 0-11             & 0-12             & 0-12             & 0-12             & 0-0              & 0-12             \\
                               & Clustering threshold   & 0.3              & NA               & NA               & 0.3              & NA               & 0.3              & NA               & NA               & NA               & NA               & NA               & 0.3              & NA               & 0.1              \\
                               & \# of neurons          & 299              & 299              & 299              & 49               & 599              & 299              & 599              & 29               & 99               & 99               & 199              & 399              & 9                & 29               \\
                               & Accuracy               & 78.28\%          & \textbf{90.05\%} & \textbf{87.05\%} & \textbf{87.84\%} & 89.26\%          & 85.64\%          & 92.75\%          & \textbf{61.38\%} & \textbf{65.15\%} & \textbf{61.46\%} & \textbf{64.60\%} & \textbf{63.65\%} & 56.55\%          & \textbf{65.56\%} \\
                               & Diff.                  & 3.67\%           & 1.42\%           & 1.61\%           & 7.59\%           & 0.88\%           & -0.29\%          & 8.73\%           & -0.08\%          & 1.13\%           & 0.19\%           & 1.02\%           & -0.22\%          & \cellcolor{Blue} 8.16\%           & -1.24\%          \\
                               & Neuron reduction       & 97.01\%          & \textbf{97.01\%} & \textbf{97.01\%} & \textbf{99.51\%} & 94.00\%          & 97.01\%          & 94.00\%          & \textbf{99.71\%} & \textbf{99.01\%} & \textbf{99.01\%} & \textbf{98.01\%} & \textbf{96.00\%} & 99.91\%          & \textbf{99.71\%} \\ \hline
\end{tabular}
\end{adjustbox}
\begin{tablenotes}
\small
\item \textbf{CB:} CodeBERT, \textbf{CGJ:} CodeGPTJava, \textbf{UC:} UniXCoder \textbf{CGP:} CodeGPTPy, \textbf{GCB:} GraphCodeBERT,\textbf{RBa:} RoBERTa,\textbf{UC:} UniXCoder.Bold numbers indicate the best score after significant neuron reduction. We highlight examples showing a significant increase in accuracy after removing redundant neurons in blue. Yellow highlights indicate a clustering threshold lower than 0.5. (Referenced in text)
\end{tablenotes}
\end{table*}

\begin{table*}[]

\definecolor{lightyellow}{RGB}{255, 255, 204}
\caption{Neuron level analysis - continued} \label{tab: Result2}
\begin{adjustbox}{width=1\textwidth}
\begin{tabular}{|c|c|ccccccc|ccccccc|}
\hline
                               &                        & \multicolumn{7}{c|}{Clone Detection}                                                                                               & \multicolumn{7}{c|}{Code Search}                                                                                                   \\ \hline
Selection                      &                        & BERT             & CB               & CGJ              & CGP              & GCB              & RBa              & UC               & BERT             & CB               & CGJ              & CGP              & GCB              & RBa              & UC               \\ \hline
\multirow{2}{*}{Oracle}        & original \# of neurons & 9984             & 9984             & 9984             & 9984             & 9984             & 9984             & 9984             & 9984             & 9984             & 9984             & 9984             & 9984             & 9984             & 9984             \\
                               & Baseline score      & 0.723            & 0.805            & 0.778            & 0.751            & 0.805            & 0.815            & 0.828            & 50.17\%          & 54.14\%          & 52.15\%          & 47.68\%          & 46.52\%          & 49.17\%          & 50.99\%          \\ \hline
\multirow{4}{*}{LCA}           & \# of neurons          & 169              & 9                & 9                & 9                & 9                & 469              & 69               & 718              & 569              & 1497             & 1497             & 429              & 539              & 599              \\
                               & Score               & \textbf{0.715}   & \textbf{0.799}   & 0.764            & 0.721            & 0.809            & 0.820            & 0.821            & 49.18\%          & 51.82\%          & 50.33\%          & 50.33\%          & 52.32\%          & 50.83\%          & 50.33\%          \\
                               & Diff.                  & -0.008           & -0.006           & -0.014           & -0.030           & 0.004            & 0.005            & -0.007           & -0.99\%          & -2.32\%          & -1.82\%          & \cellcolor{Blue}2.65\%           & \cellcolor{Blue}5.80\%           & \cellcolor{Blue}1.66\%           & -0.66\%          \\
                               & Neuron reduction       & \textbf{98.31\%} & \textbf{99.91\%} & 99.91\%          & 99.91\%          & 99.91\%          & 95.30\%          & 99.31\%          & 92.81\%          & 94.30\%          & 85.01\%          & 85.01\%          & 95.70\%          & 94.60\%          & 94.00\%          \\ \hline
\multirow{5}{*}{CC}            & Clustering threshold   & \cellcolor{lightyellow} 0.30             & 0.6              & \cellcolor{lightyellow} 0.3              & \cellcolor{lightyellow}0.2              & \cellcolor{lightyellow}0.4              & 0.6              & 0.8              & 0.7              & 0.9              & 0.5              & 0.8              & 0.5              & 0.6              & 0.8              \\
                               & \# of neurons          & 3121             & 2264             & 1679             & 1416             & 6124             & 1110             & 901              & 768              & 768              & 2303             & 1536             & 1454             & 768              & 768              \\
                               & Score               & 0.727            & 0.823            & 0.811            & 0.814            & 0.801            & 0.802            & 0.805            & 52.48\%          & 55.46\%          & 52.15\%          & 49.34\%          & 53.48\%          & 53.31\%          & 52.32\%          \\
                               & Diff.                  & 0.004            & 0.018            & 0.033            & 0.063            & -0.004           & -0.013           & -0.023           & \cellcolor{Blue}2.31\%           & 1.32\%           & 0.00\%           & \cellcolor{Blue}1.66\%           & \cellcolor{Blue}6.96\%           & \cellcolor{Blue}4.14\%           & 1.33\%           \\
                               & Neuron reduction       & 68.74\%          & 77.32\%          & 83.18\%          & 85.82\%          & 38.66\%          & 88.88\%          & 90.98\%          & 92.31\%          & 92.31\%          & 76.93\%          & 84.62\%          & 85.44\%          & 92.31\%          & 92.31\%          \\ \hline
\multirow{5}{*}{Layerwise(LS)} & Layer Selection        & 0-12             & 0-11             & 0-11             & 0-11             & 0-11             & 0-12             & 0-12             & 0-9              & 0-10             & 0-11             & 0-2              & 0-1              & 0-1              & 0-11             \\
                               & \# of neurons          & 9984             & 9216             & 9216             & 9216             & 9216             & 9984             & 9984             & 7680             & 8448             & 9216             & 2304             & 1536             & 1536             & 9216             \\
                               & Score               & 0.723            & 0.819            & 0.783            & 0.824            & 0.807            & 0.815            & 0.828            & 51.66\%          & 54.64\%          & 52.32\%          & 49.50\%          & 52.15\%          & 52.15\%          & 52.81\%          \\
                               & Diff.                  & 0.000            & 0.014            & 0.005            & 0.073            & 0.002            & 0.000            & 0.000            & 1.49\%           & 0.50\%           & 0.17\%           & \cellcolor{Blue}1.82\%           & \cellcolor{Blue}5.63\%           & \cellcolor{Blue}2.98\%           & 1.82\%           \\
                               & Neuron reduction       & 0.00\%           & 7.69\%           & 7.69\%           & 7.69\%           & 7.69\%           & 0.00\%           & 0.00\%           & 23.08\%          & 15.38\%          & 7.69\%           & 76.92\%          & 84.62\%          & 84.62\%          & 7.69\%           \\ \hline
\multirow{6}{*}{LS+CC+LCA}     & Layer Selection        & 0-12             & 0-7              & 0-11             & 0-2              & 0-11             & 0-6              & 0-12             & 0-1              & 0-1              & 0-4              & 0-3              & 0-1              & 0-1              & 0-1              \\
                               & Clustering threshold   & 0.1              & 0.4              & 0.1              & NA               & NA               & NA               & NA               & 0.9              & 0.9              & 0.8              & 0.9              & 0.1              & 0.1              & 0.1              \\
                               & \# of neurons          & 299              & 19               & 199              & 49               & 19               & 9                & 99               & 399              & 399              & 499              & 399              & 399              & 499              & 499              \\
                               & Score               & 0.714            & 0.798            & \textbf{0.828}   & \textbf{0.781}   & \textbf{0.814}   & \textbf{0.835}   & \textbf{0.831}   & \textbf{51.49\%} & \textbf{53.81\%} & \textbf{53.97\%} & \textbf{53.15\%} & \textbf{53.31\%} & \textbf{53.97\%} & \textbf{52.32\%} \\
                               & Diff.                  & -0.009           & -0.007           & 0.050            & 0.030            & 0.009            & 0.020            & 0.003            & 1.32\%           & -0.33\%          & 1.82\%           & \cellcolor{Blue}5.47\%           & \cellcolor{Blue}6.79\%           & \cellcolor{Blue}4.80\%           & 1.33\%           \\
                               & Neuron reduction       & 97.01\%          & 99.81\%          & \textbf{98.01\%} & \textbf{99.51\%} & \textbf{99.81\%} & \textbf{99.91\%} & \textbf{99.01\%} & \textbf{96.00\%} & \textbf{96.00\%} & \textbf{95.00\%} & \textbf{96.00\%} & \textbf{96.00\%} & \textbf{95.00\%} & \textbf{95.00\%} \\ \hline
\end{tabular}
\end{adjustbox}
\begin{tablenotes}
\small
\item Score: F1 score for Clone Detection, accuracy for Code Search
\end{tablenotes}
\end{table*}

\subsubsection{Redundancy in shared encoding space} 

Despite a shared encoding space, we find that in models trained across multiple programming languages like CodeBERT, GraphCodeBERT, multiple pretraining objectives or different modalities like UniXCoder, GraphCodeBERT, etc. (table \ref{tab:Models}) \textbf{more than 95\% on neurons can be eliminated due to redundancy} without losing probing accuracy (table \ref{tab: Results1} and \ref{tab: Result2}). Therefore, a very small number of neurons is required to encode information about all our software engineering tasks.


\subsubsection{Increase in accuracy upon eliminating redundant neurons}

We observe that, in several cases, there is an increase in accuracy after eliminating redundant neurons using different neuron selection techniques. We highlight cases with an increase in accuracy greater than 1.5\% in blue in the result tables. We note that certain models show an increase across all neuron selection techniques. For example, \textbf{in the case of defect detection (Table \ref{tab: Results1}), RoBERTa (RBa) exhibits an accuracy increase of more than 5\% after reducing neurons using every neuron selection technique. In codesearch (Table \ref{tab: Result2}), RoBERTa (RBa), CodeGPT-Python (CGP), and GraphCodeBERT (GCB) see increased accuracies upon eliminating more than 76\% of neurons.} This increased accuracy could indicate that there is noise in the network concerning the given task, and eliminating specific neurons reduces this noise, leading to better accuracies. However, this could also suggest a reliance on spurious features when using the reduced set of neurons, which merits further investigation. Additionally, we observe that clone detection models do not see an increase in F1 score (Table \ref{tab: Result2}) upon eliminating redundancies, which could indicate that the entire activation space makes use of all important information and is not affected by noise.

\subsubsection{Adjacent layers are most redundant due to similarity}
\label{Evaluation-Redundancy Analysis-Adjacent layers are most redundant due to similarity}

We performed Centered Kernel Alignment (CKA) to measure how similar the layers of our models are to each other. We find that similar to NLP \citep{dalvi2020analyzing}, for most of our models, adjacent layers are similar to each other (Fig. \ref{fig:cka_token}). Further, \textbf{in the case of CodeGPT-Java and CodeGPT-Py models for the token-tagging task, layers 3 to 11 are highly similar for both models (Figure \ref{fig:cka_token}) and can be considered redundant.} We believe this is because these two models were only trained on monolingual pretraining data and therefore have less information encoded within them. We also observe that the CKA results for each of the different fine-tuned models are different from the pretrained models. For example, the layers of the CodeBERT model for the Clone Detection tasks are highly dissimilar, while they are quite similar for the defect detection task. \textbf{Overall, code search task models seem to have the most dissimilar adjacent layers (Figure \ref{fig:cka_token}).} Further, the early layers of the CodeGPT-Java and CodeGPT-Py models for Code Search are not similar, but the later layers are. For several multilingual models, we observe that the later layers are more similar, suggesting that there is not much new information encoded within them. It is well-known that the later layers tend to focus on higher-level concepts. The similar layers can be pruned if those concepts are learned in lower layers.

\subsubsection{Influence of additional modalities like code structure and natural language}

GraphCodeBERT and UniXCoder are two models that have been pretrained with additional code structure information like data flow graphs or abstract syntax trees. While we do not generally see a marked difference in probing performances for models with code structure information across tasks, it might influence some tasks. Correlation clustering performed on the UniXCoder model for token tagging does not allow any reduction in neurons, while all other models see a reduction greater than 70\%. We also see a decreased neuron reduction percentage in GraphCodeBERT (74.41\%) and UniXCoder (10.96\%) on the defect detection task. In clone detection, GraphCodeBERT correlation clustering neuron reduction is only 38\%. This could suggest that token tagging information is more distributed because the activations are updated due to multiple training objectives and modalities. \textbf{We do not see an overarching trend for models pretrained on natural language.}

\subsubsection{Influence of model architecture}
Encoder-based models are generally used for code understanding tasks as their bidirectional nature can help the model learn better contextual embeddings. \textbf{However, we do not see a discernable general trend that suggests that decoder-based generative models do not encode relevant information about the tasks.} Whether that information can be accessed or used needs to be investigated through further studies. 


\subsubsection{Implications in knowledge distillation and model compression}
\label{Evaluation- Implicaitons model distillation and pruning}
While several application-specific conclusions can be made from our data, we discuss some trends that we observed.  Models that can learn more specialized concepts in the lower layers, like GraphCodeBERT, which selects layers 0-1 for code search or RoBERTa which only needs layers 0-5 for defect detection, can be pruned in the higher layers and hence make better candidates for model compression. Similarly, in cases where higher layers are able to perform well independently, they can also be pruned to improve inference costs. We share our layerwise graphs in the supplementary material, which demonstrate such cases, but we do not see clear trends from those graphs.

By studying the concentration and distribution of important neurons for different models and tasks, we obtain some insight into how decomposable and transferrable task-relevant information is. If important neurons are more concentrated, it is easier to decompose the network into modules for inference or combine them with neurons optimized for other tasks without affecting the activations of the current task. This requires a better understanding of the distribution of important concepts within the network.

 \begin{figure}
\centering
\begin{subfigure}{.12\textwidth}
  \centering
  \includegraphics[width=1\linewidth]{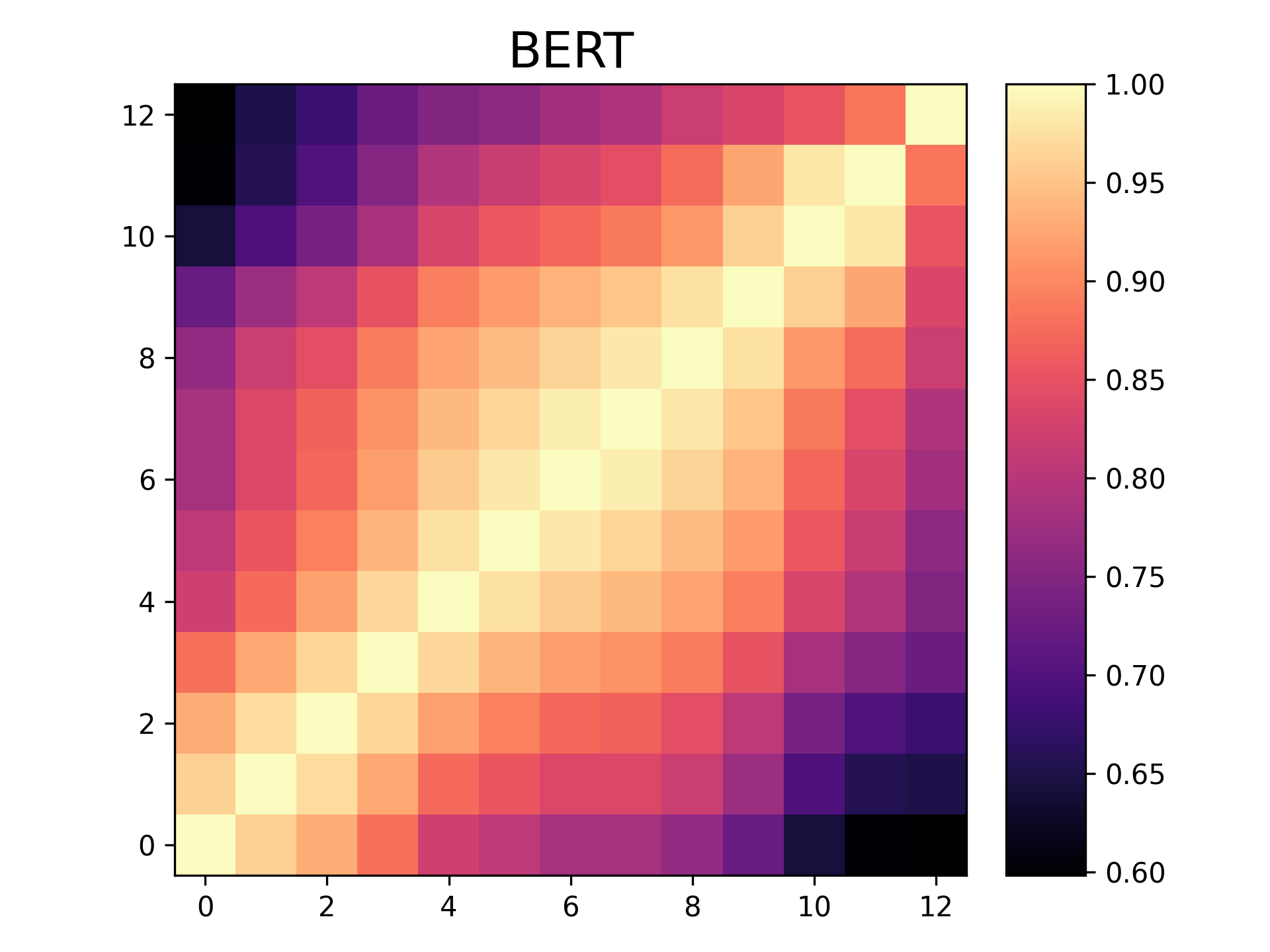}
  \label{fig:bert_cka}
\end{subfigure}%
\begin{subfigure}{.12\textwidth}
  \centering
  \includegraphics[width=1\linewidth]{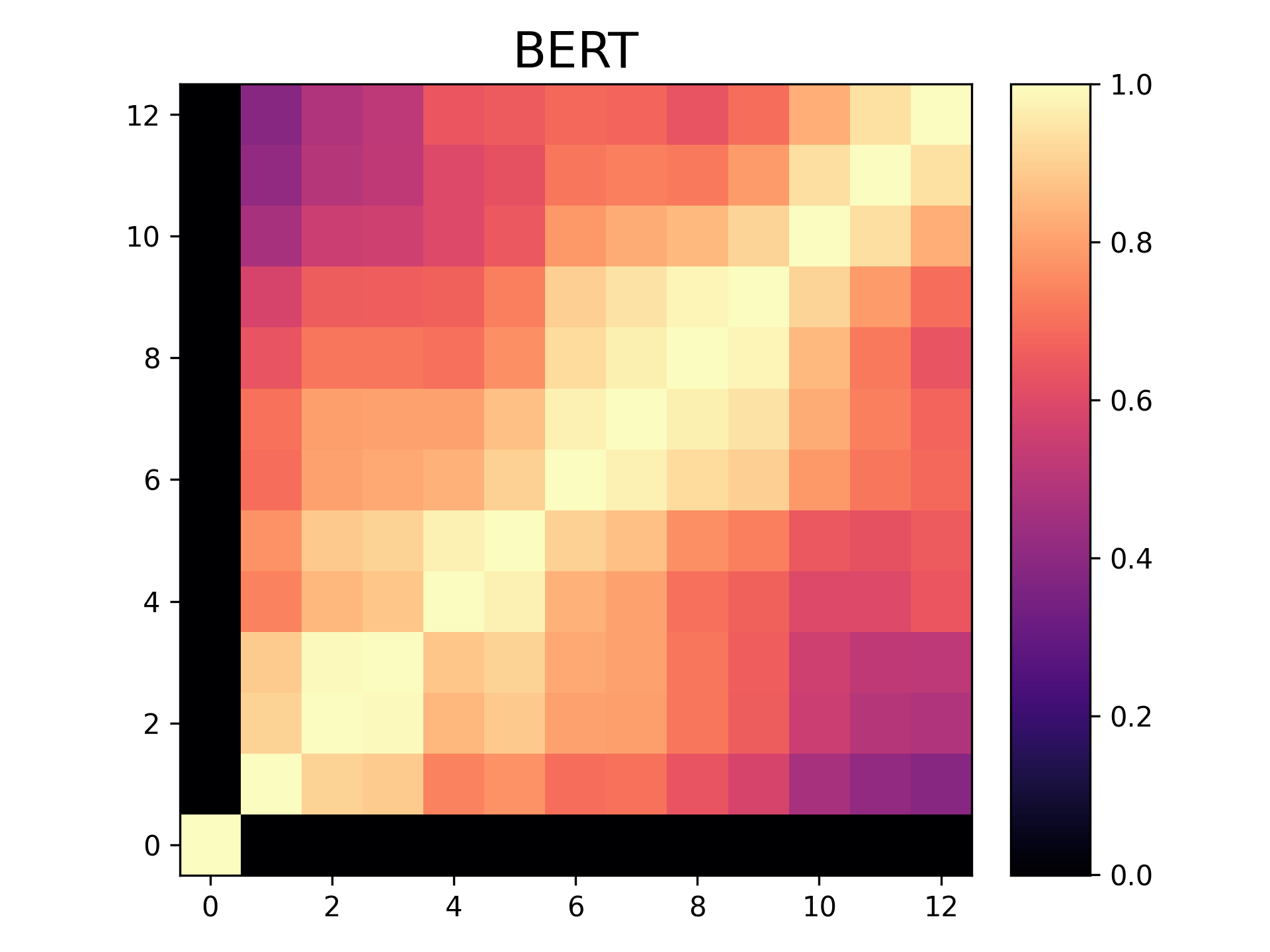}
  \label{fig:bert_cka_clonedet}
\end{subfigure}%
\begin{subfigure}{.12\textwidth}
  \centering
  \includegraphics[width=1\linewidth]{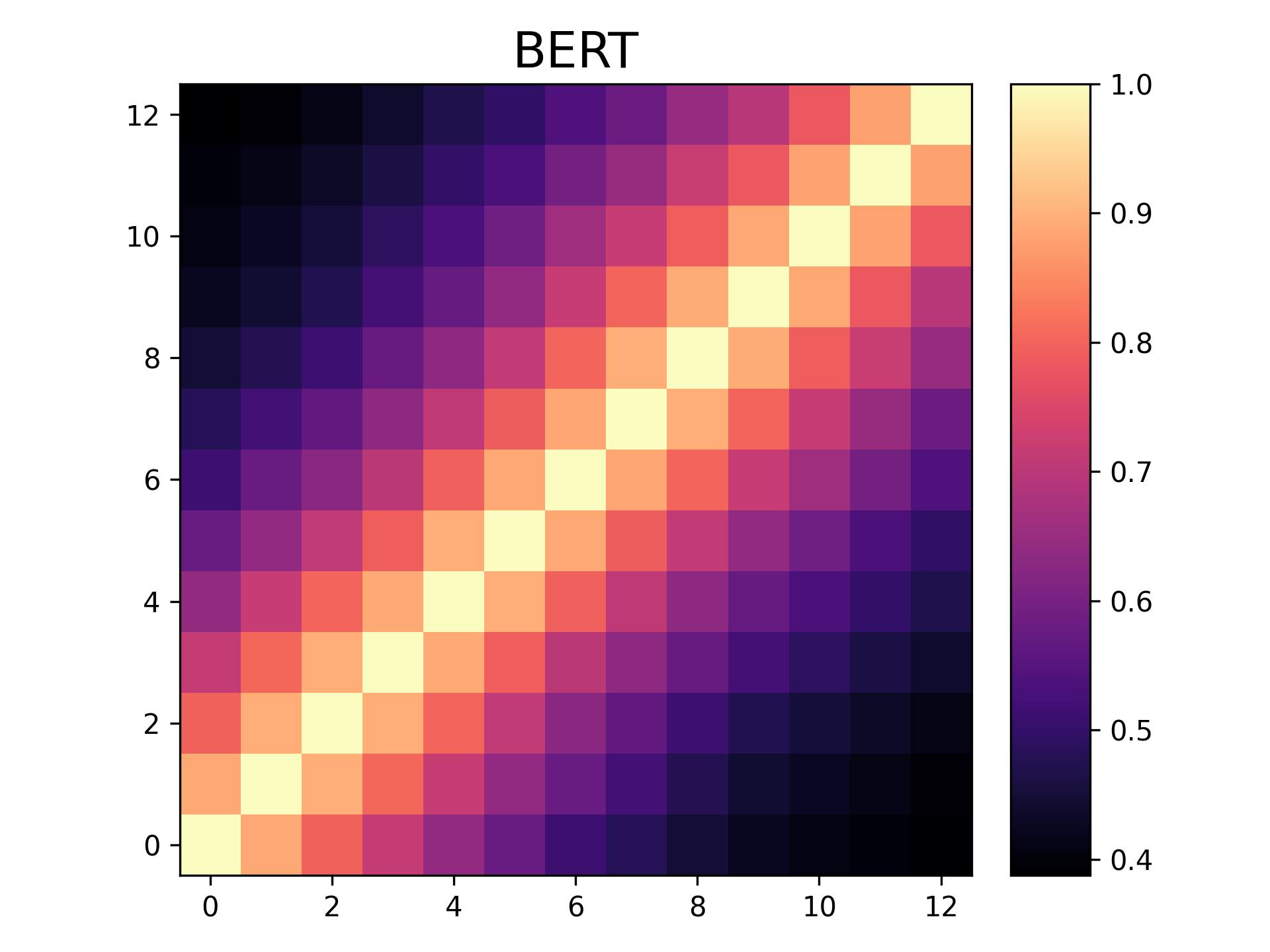}
  \label{fig:bert_cka_codesearch}
\end{subfigure}%
\begin{subfigure}{.12\textwidth}
  \centering
  \includegraphics[width=1\linewidth]{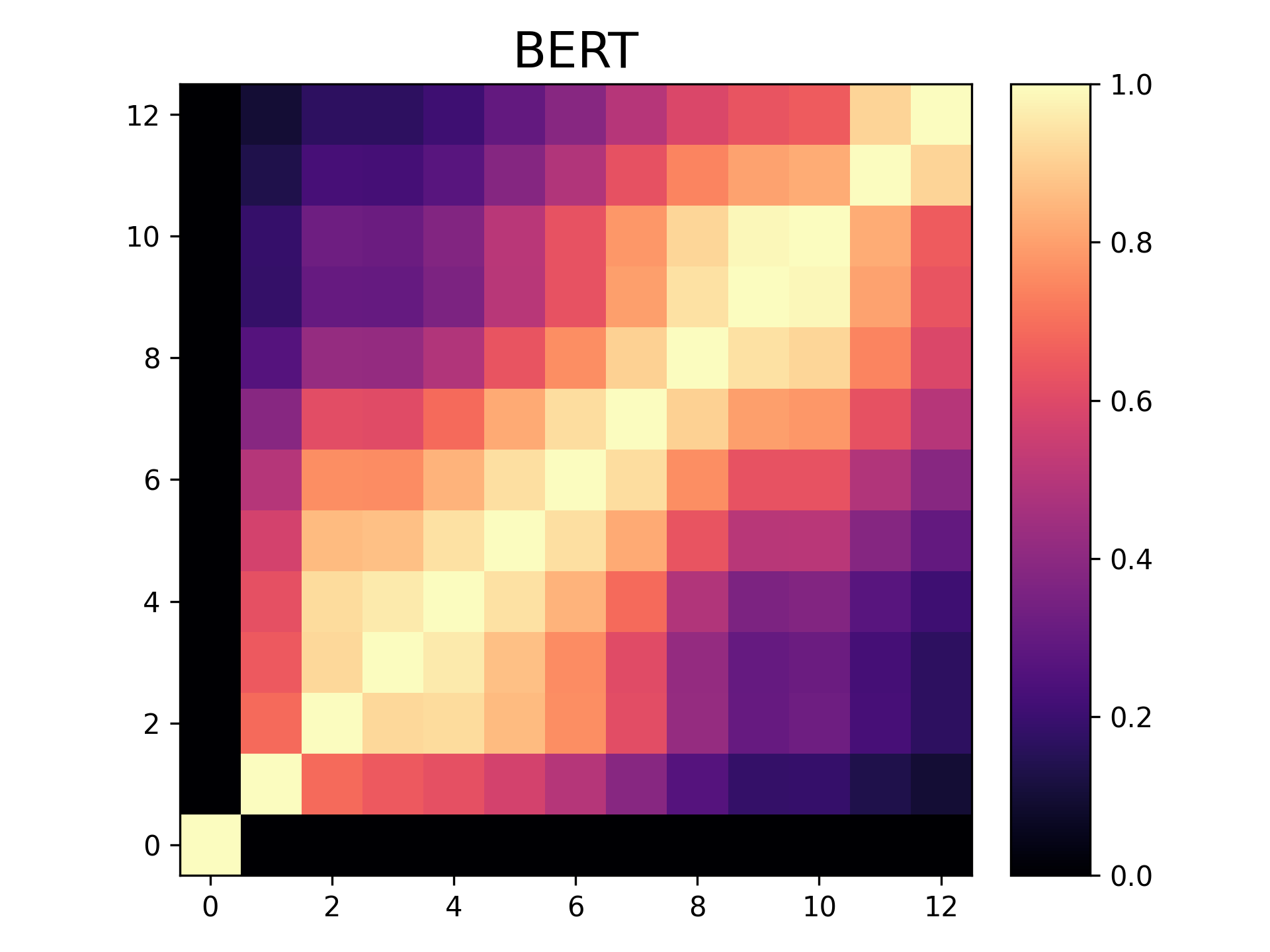}
  \label{fig:bert_cka_defdet}
\end{subfigure}%
\\
\begin{subfigure}{.12\textwidth}
  \centering
  \includegraphics[width=1\linewidth]{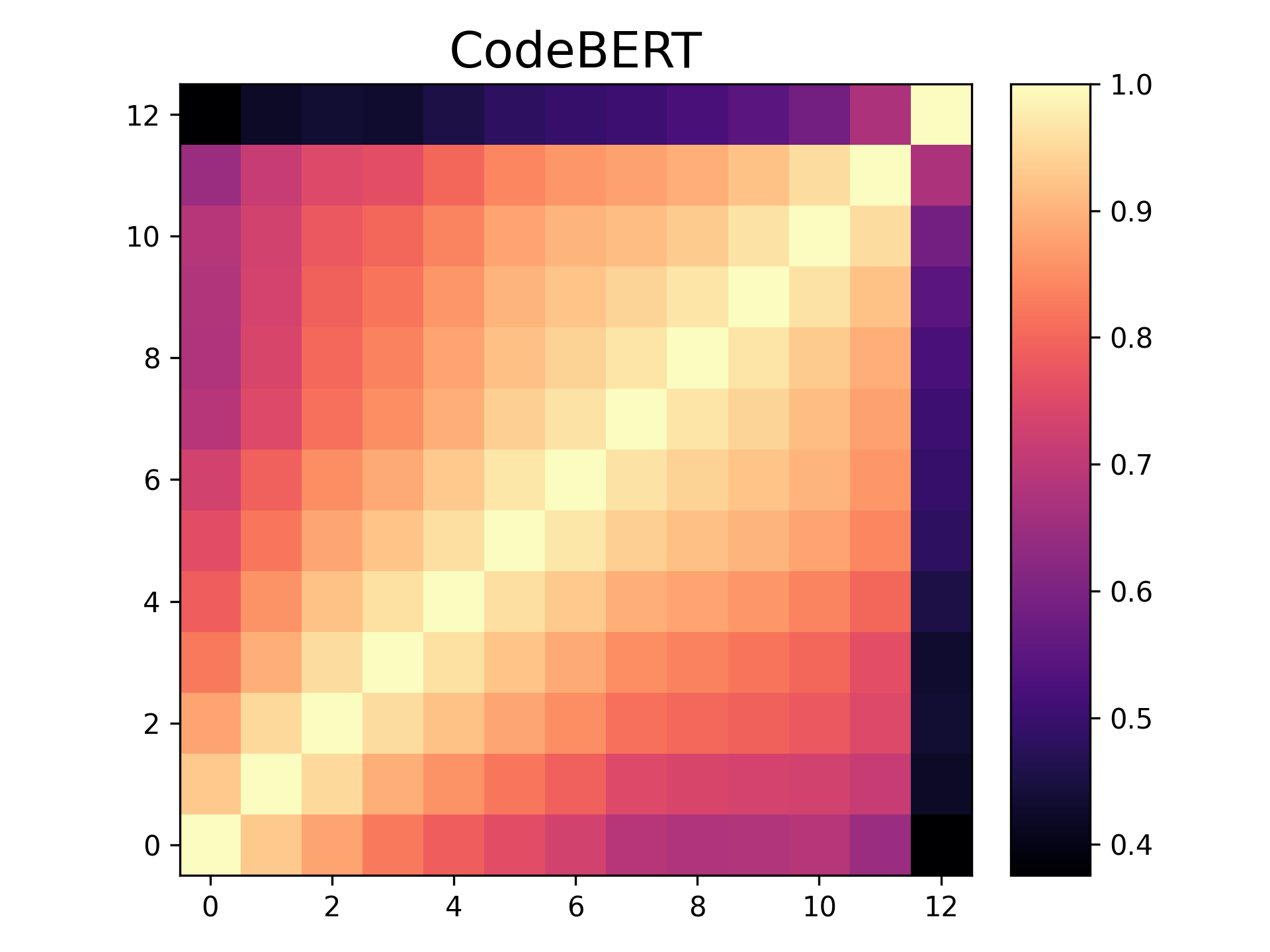}
  \label{fig:codebert_cka}
\end{subfigure}%
\begin{subfigure}{.12\textwidth}
  \centering
  \includegraphics[width=1\linewidth]{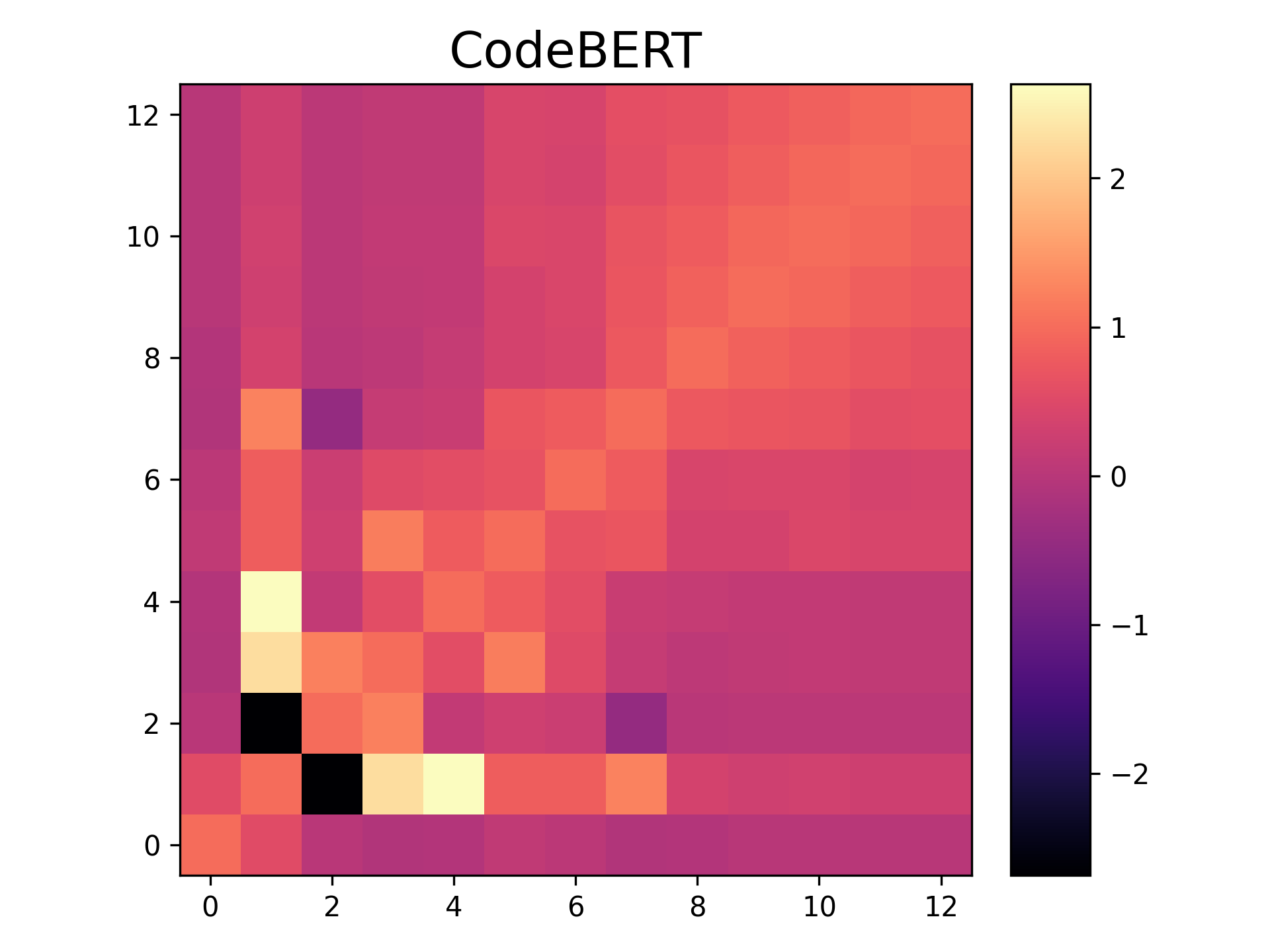}
  \label{fig:codebert_cka_clonedet}
\end{subfigure}%
\begin{subfigure}{.12\textwidth}
  \centering
  \includegraphics[width=1\linewidth]{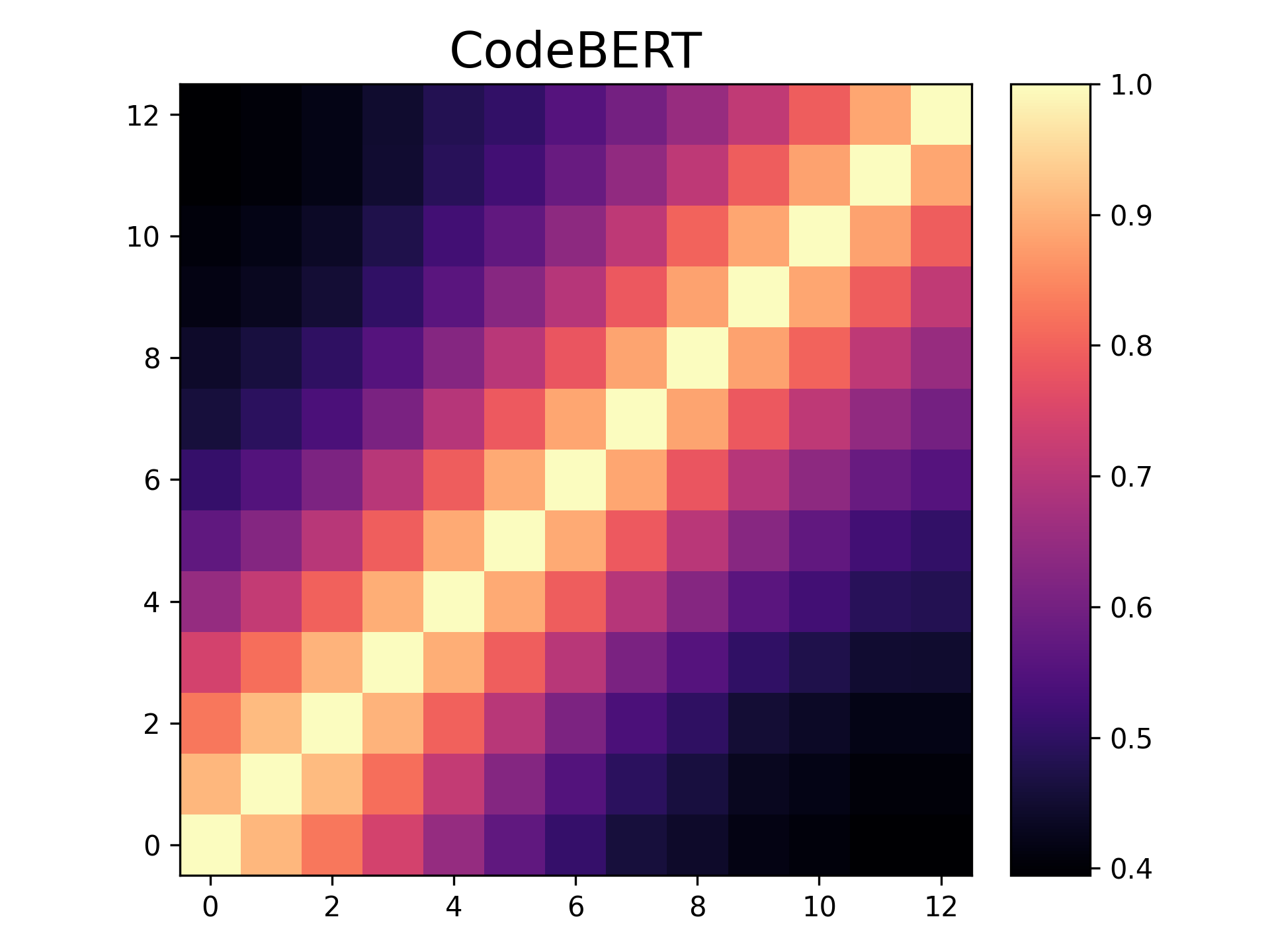}
  \label{fig:codebert_cka_codesearch}
\end{subfigure}%
\begin{subfigure}{.12\textwidth}
  \centering
  \includegraphics[width=1\linewidth]{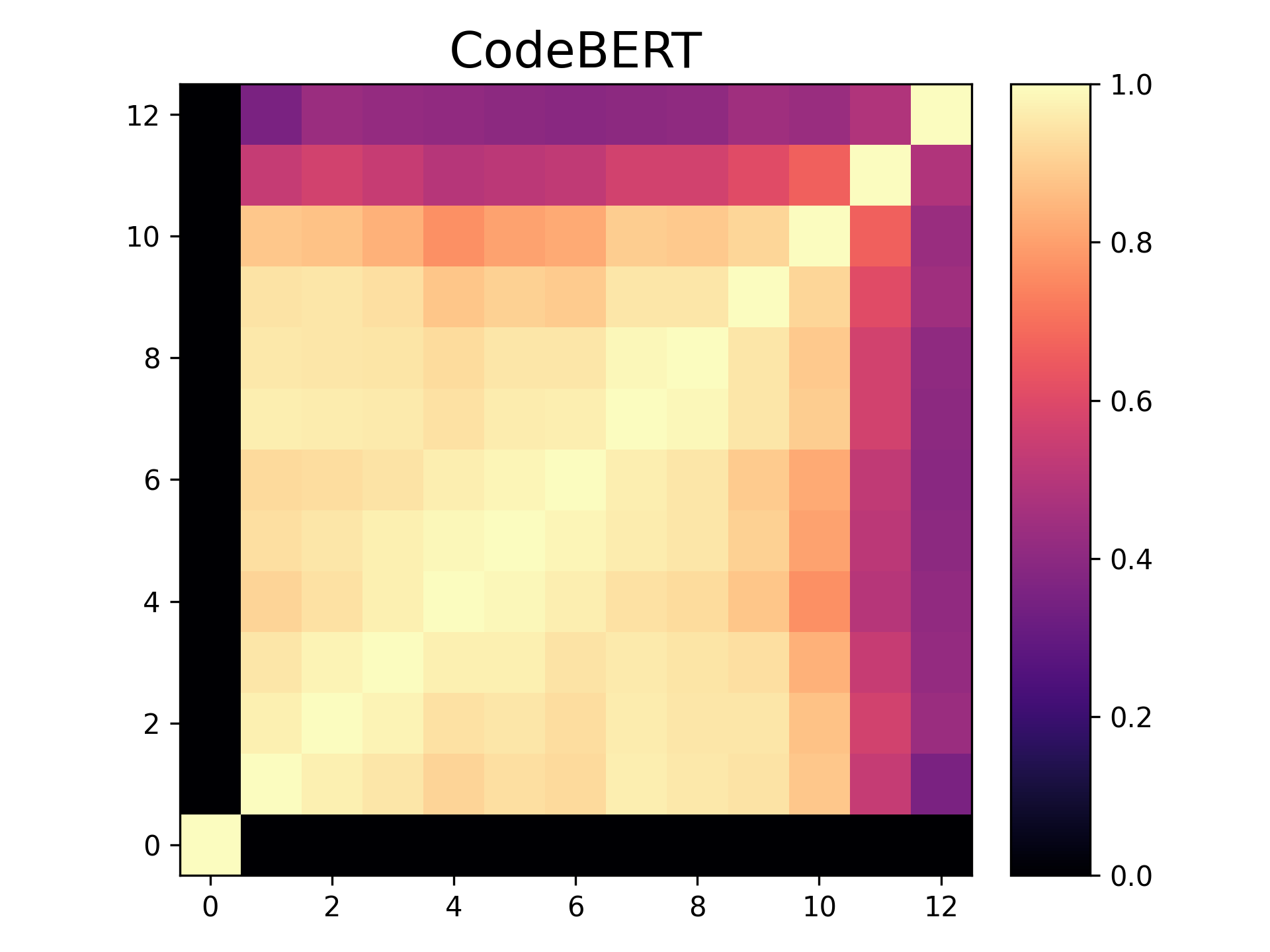}
  \label{fig:codebert_cka_defdet}
\end{subfigure}%
\\
\begin{subfigure}{.12\textwidth}
  \centering
  \includegraphics[width=1\linewidth]{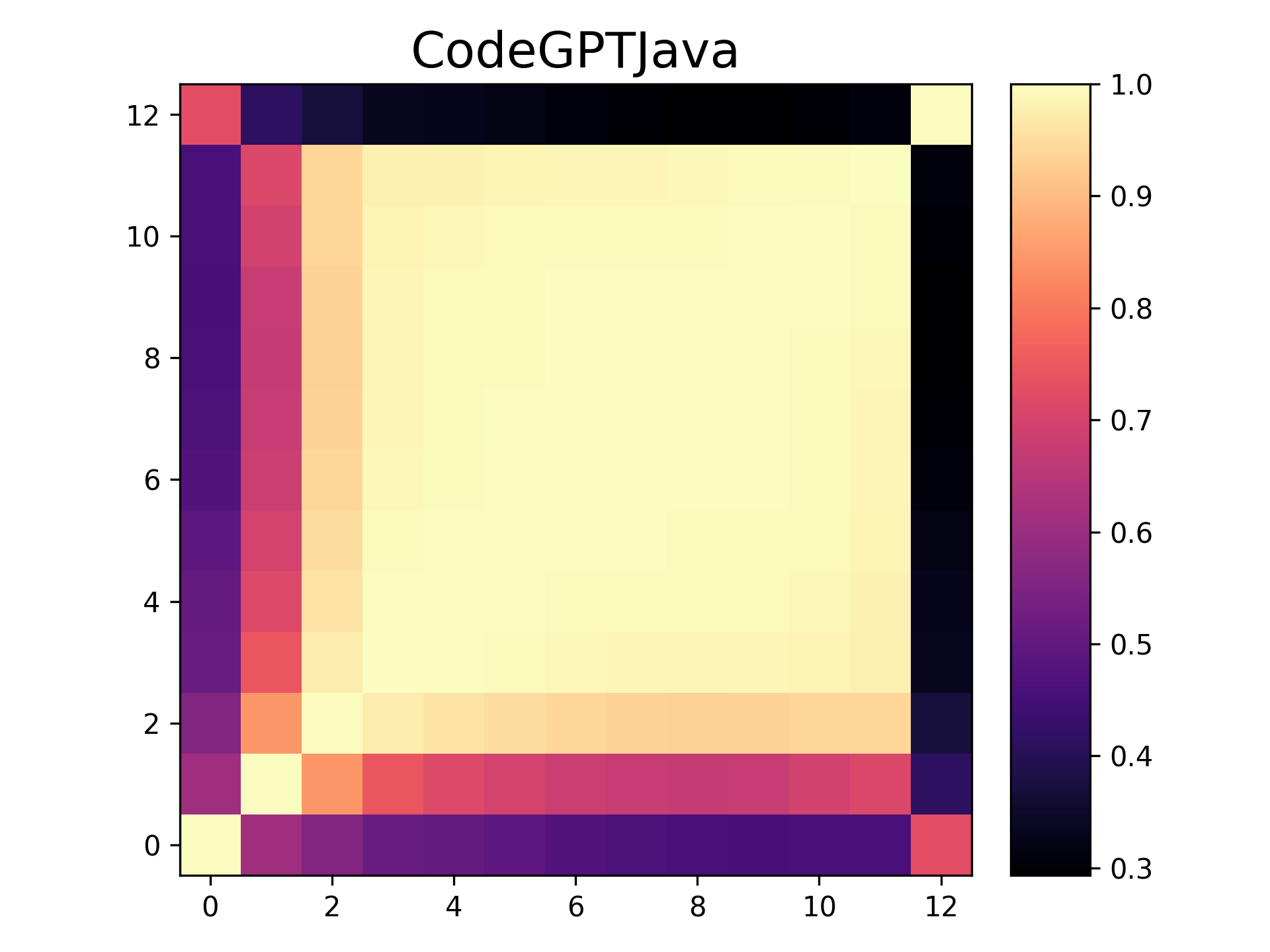}
  \label{fig:codegptjava_cka}
\end{subfigure}%
\begin{subfigure}{.12\textwidth}
  \centering
  \includegraphics[width=1\linewidth]{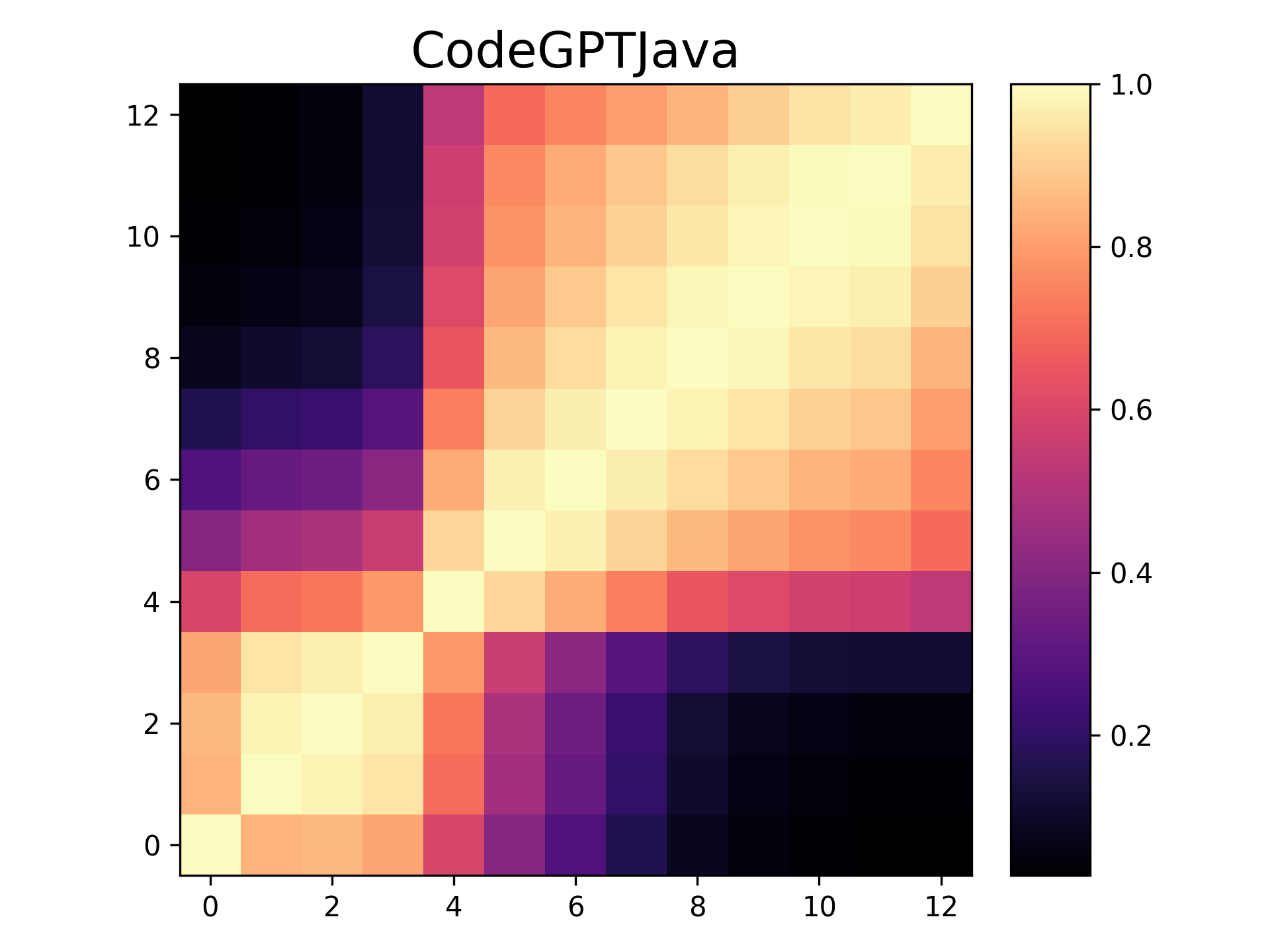}
  \label{fig:codegptjava_cka_clonedet}
\end{subfigure}%
\begin{subfigure}{.12\textwidth}
  \centering
  \includegraphics[width=1\linewidth]{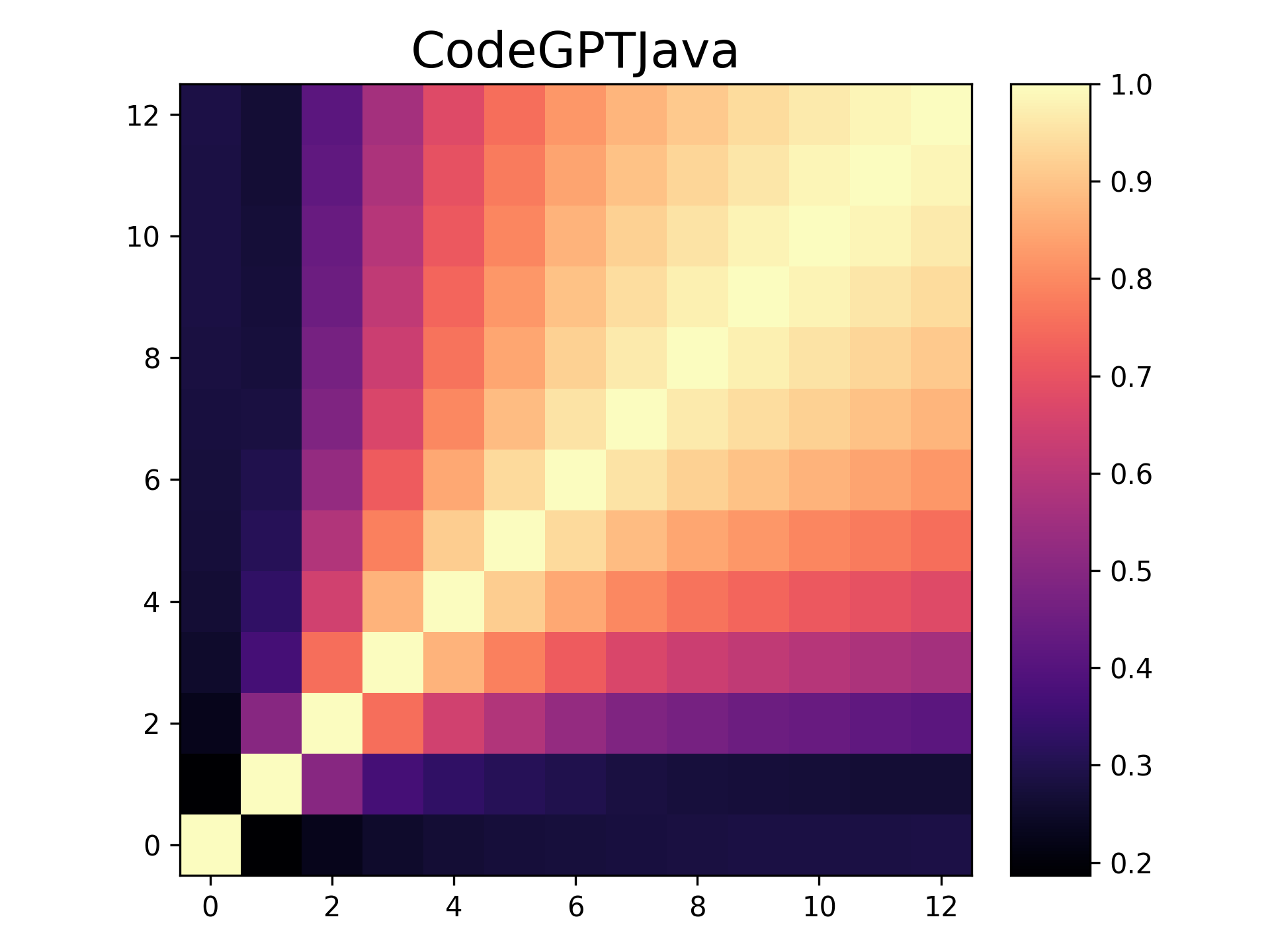}
  \label{fig:codegptjava_cka_codesearch}
\end{subfigure}%
\begin{subfigure}{.12\textwidth}
  \centering
  \includegraphics[width=1\linewidth]{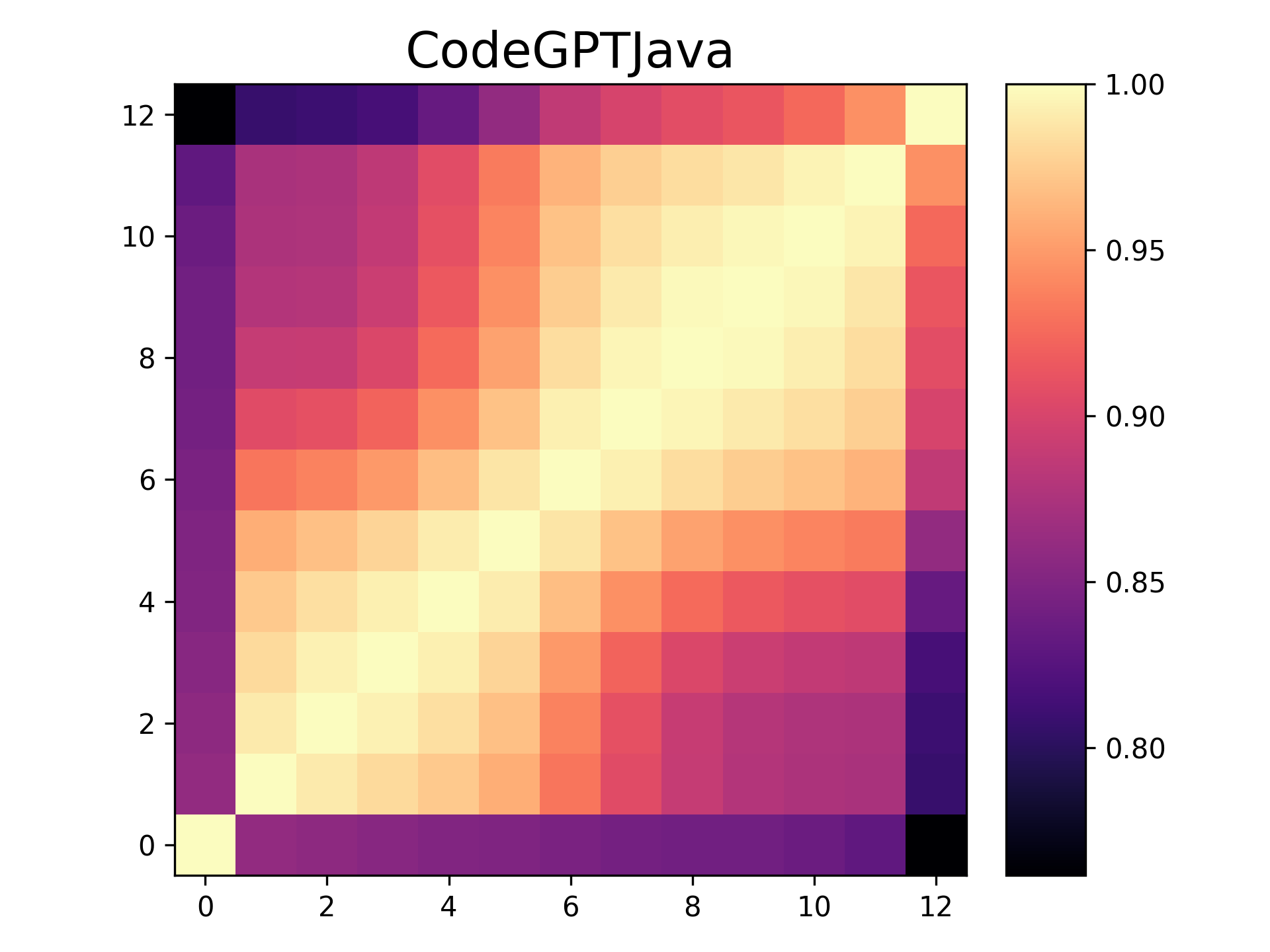}
  \label{fig:codegptjava_cka_defdet}
\end{subfigure}%
\\
\begin{subfigure}{.12\textwidth}
  \centering
  \includegraphics[width=1\linewidth]{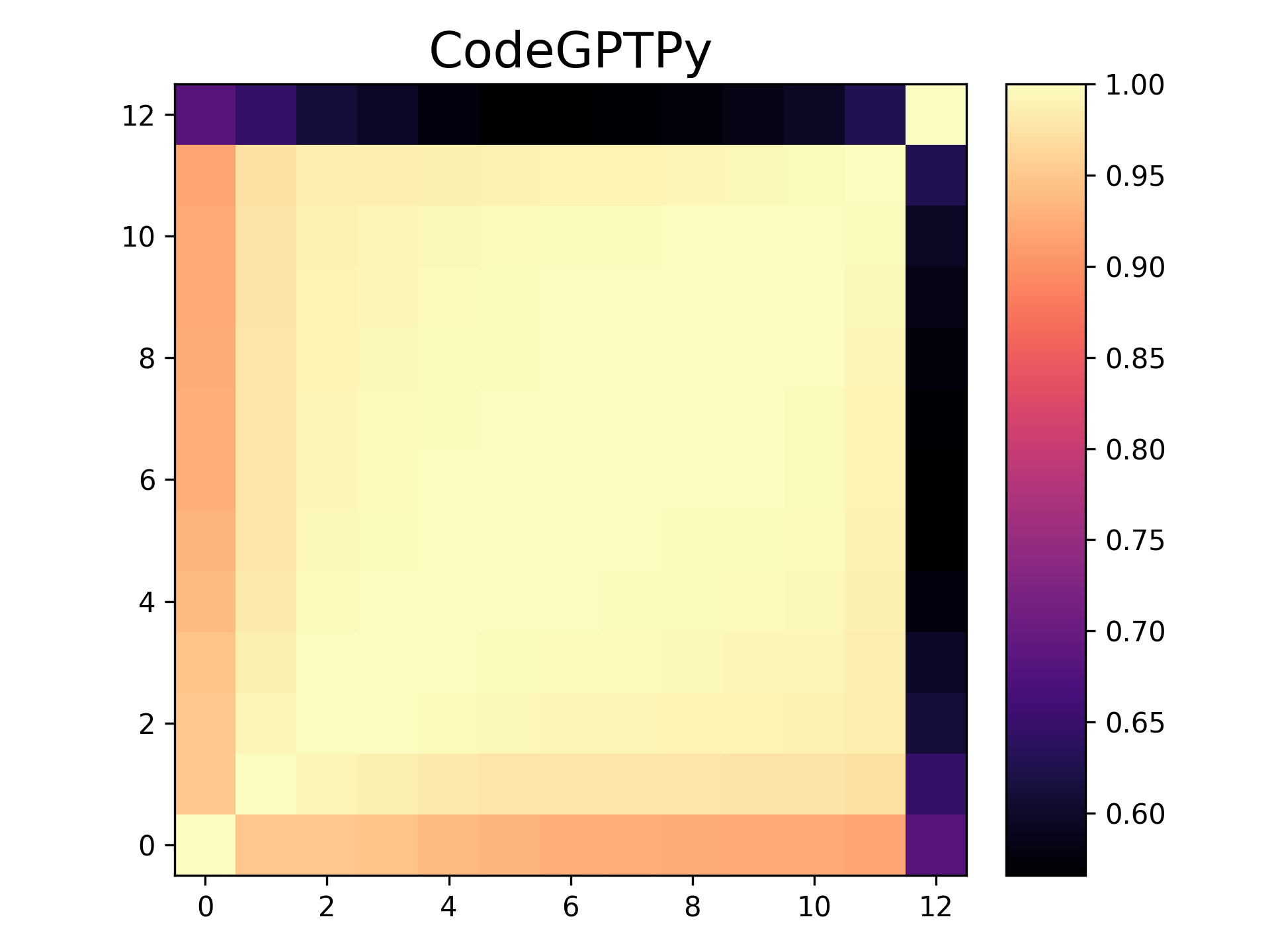}
  \label{fig:codegptpy_cka}
\end{subfigure}%
\begin{subfigure}{.12\textwidth}
  \centering
  \includegraphics[width=1\linewidth]{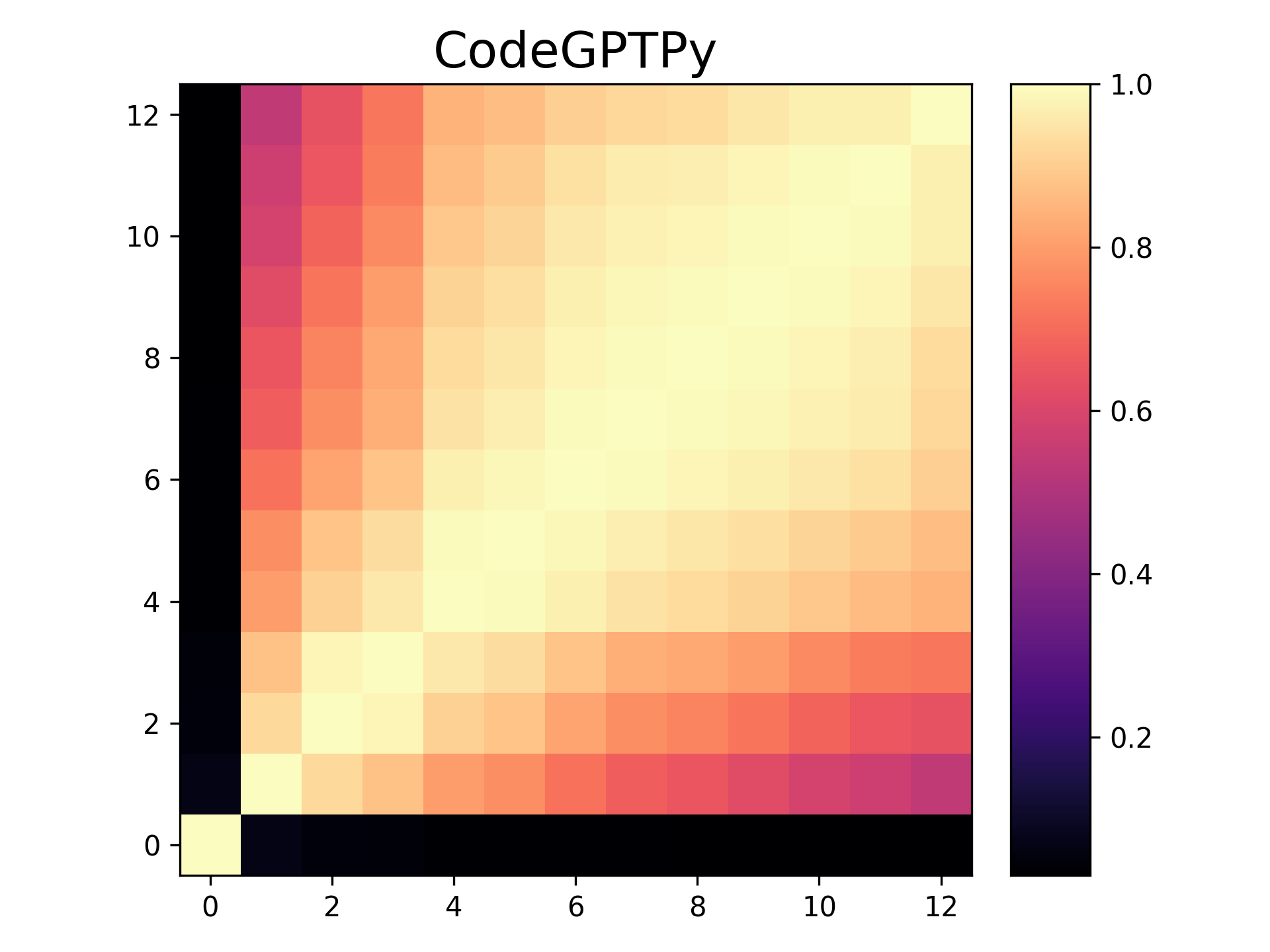}
  \label{fig:codegptpy_cka_clonedet}
\end{subfigure}%
\begin{subfigure}{.12\textwidth}
  \centering
  \includegraphics[width=1\linewidth]{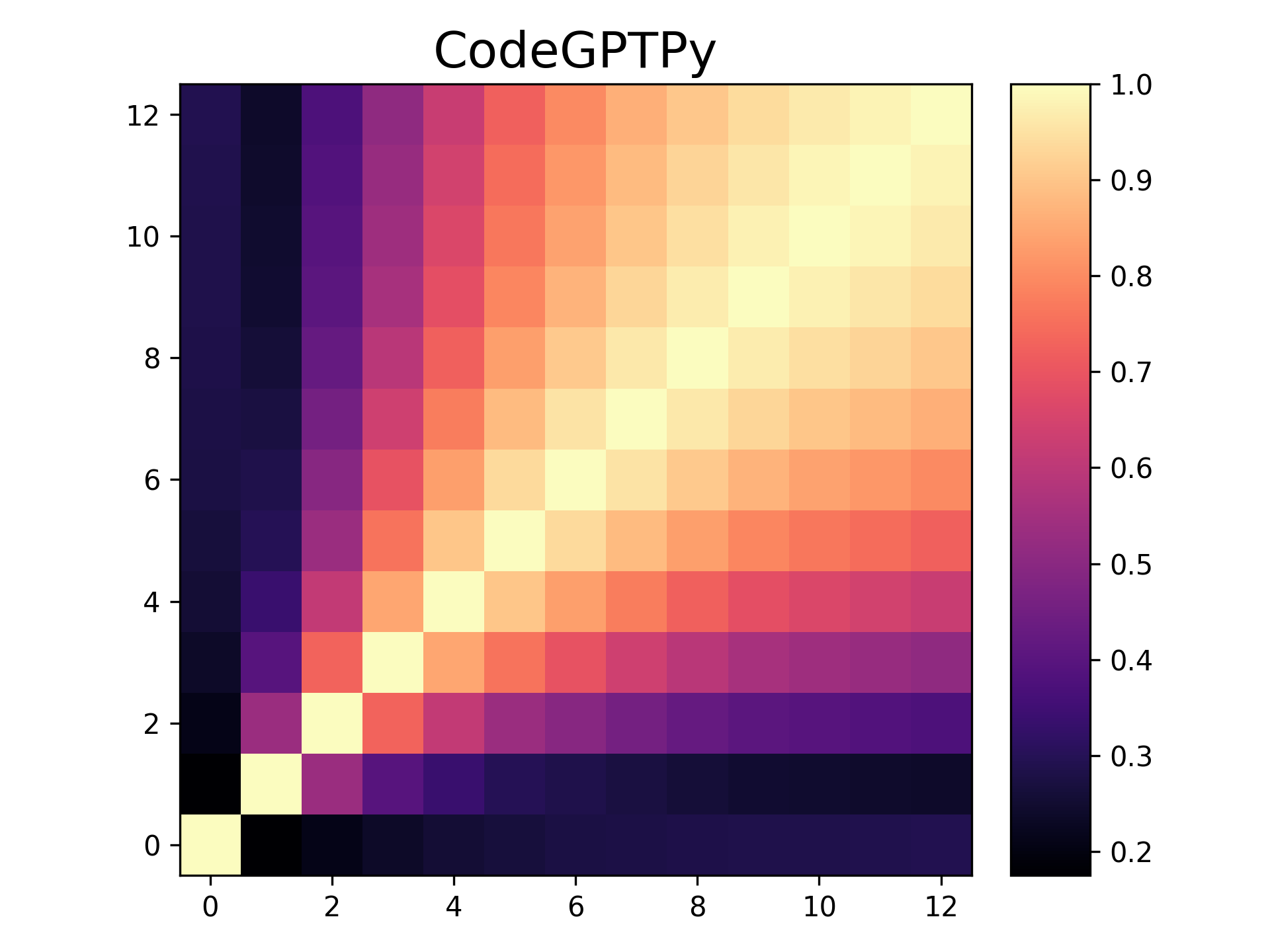}
  \label{fig:codegptpy_cka_codesearch}
\end{subfigure}%
\begin{subfigure}{.12\textwidth}
  \centering
  \includegraphics[width=1\linewidth]{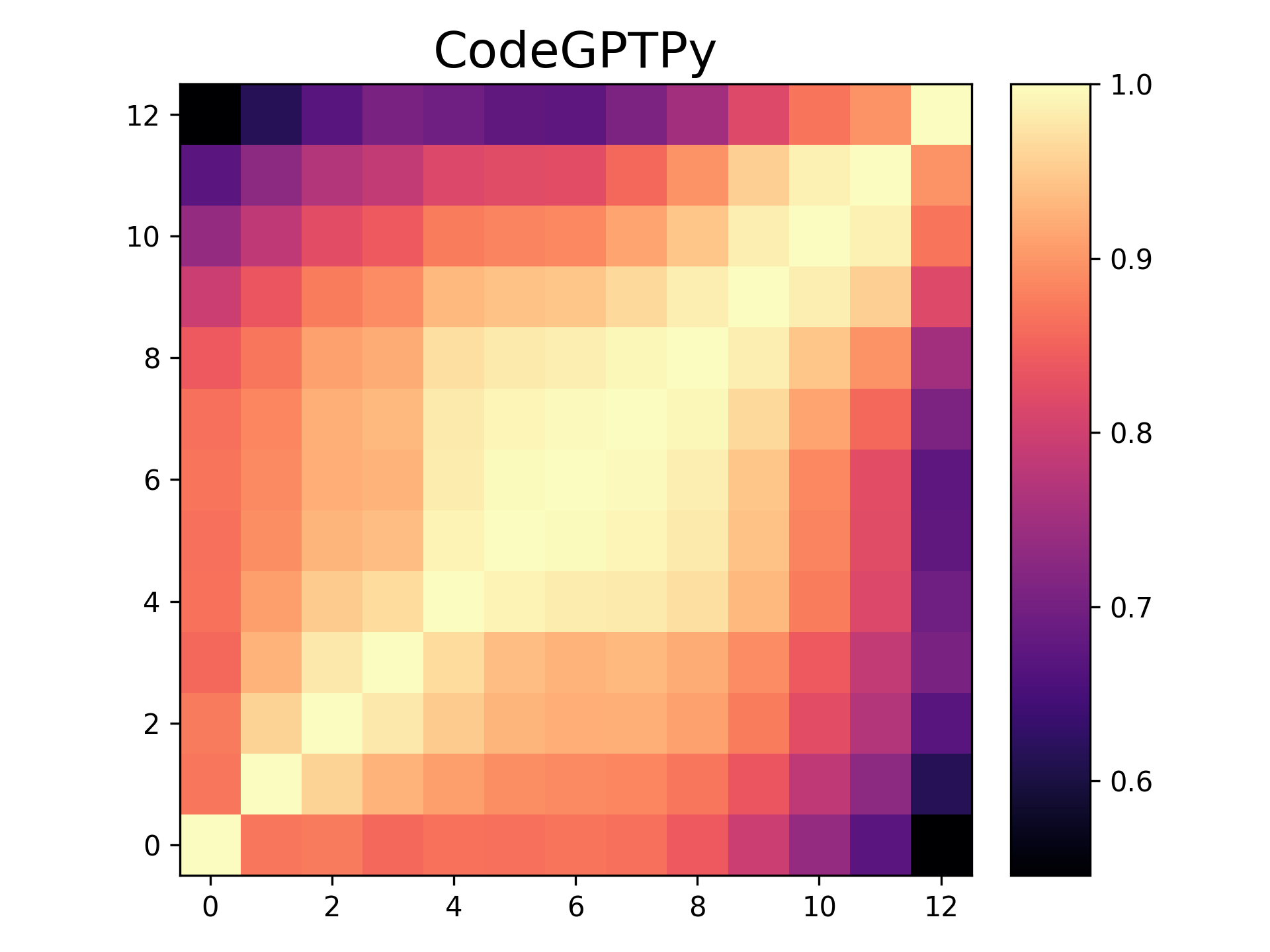}
  \label{fig:codegptpy_cka_defdet}
\end{subfigure}%
\\
\begin{subfigure}{.12\textwidth}
  \centering
  \includegraphics[width=1\linewidth]{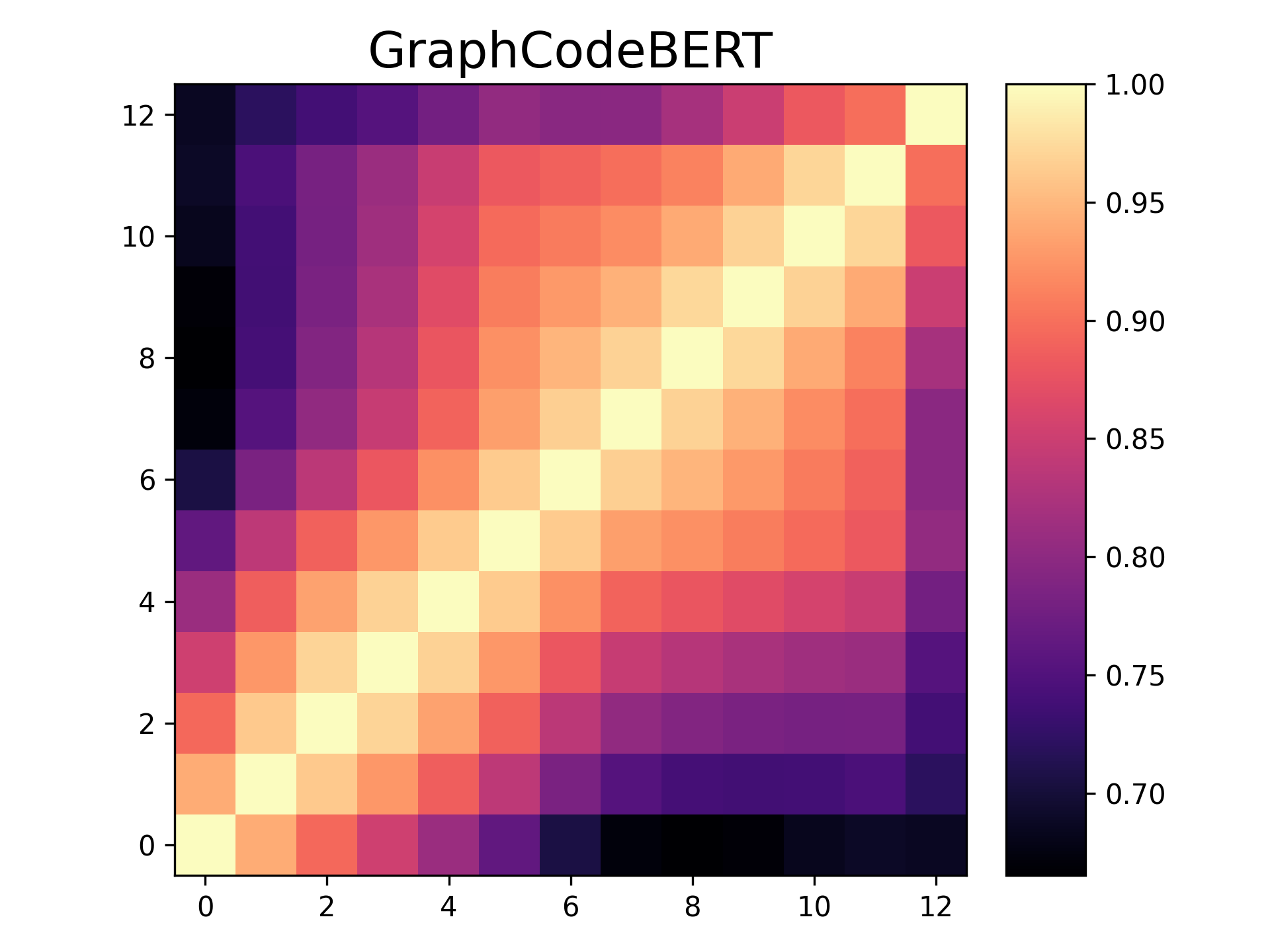}
  \label{fig:graphcodebert_cka}
\end{subfigure}%
\begin{subfigure}{.12\textwidth}
  \centering
  \includegraphics[width=1\linewidth]{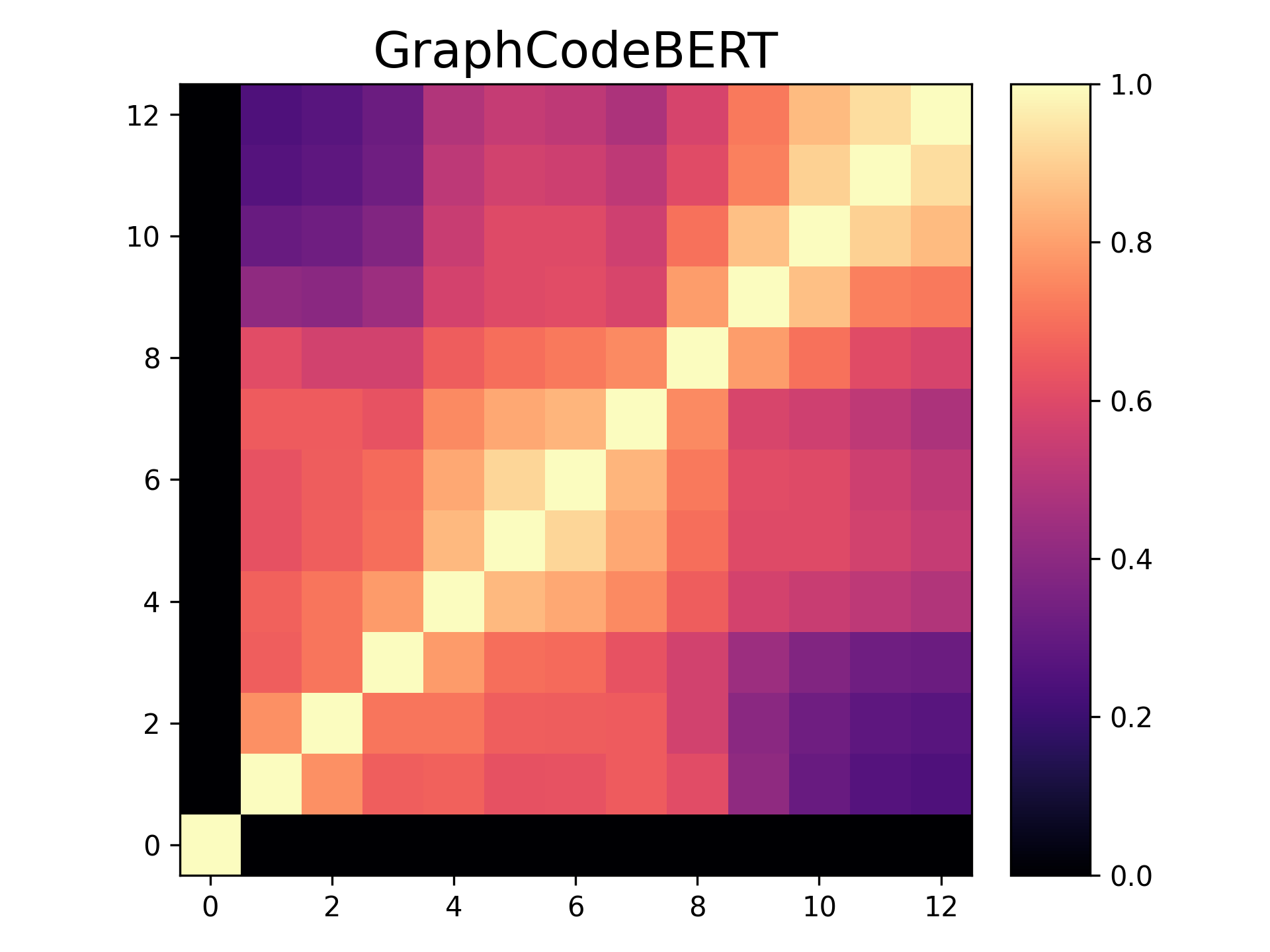}
  \label{fig:graphcodebert_cka_clonedet}
\end{subfigure}%
\begin{subfigure}{.12\textwidth}
  \centering
  \includegraphics[width=1\linewidth]{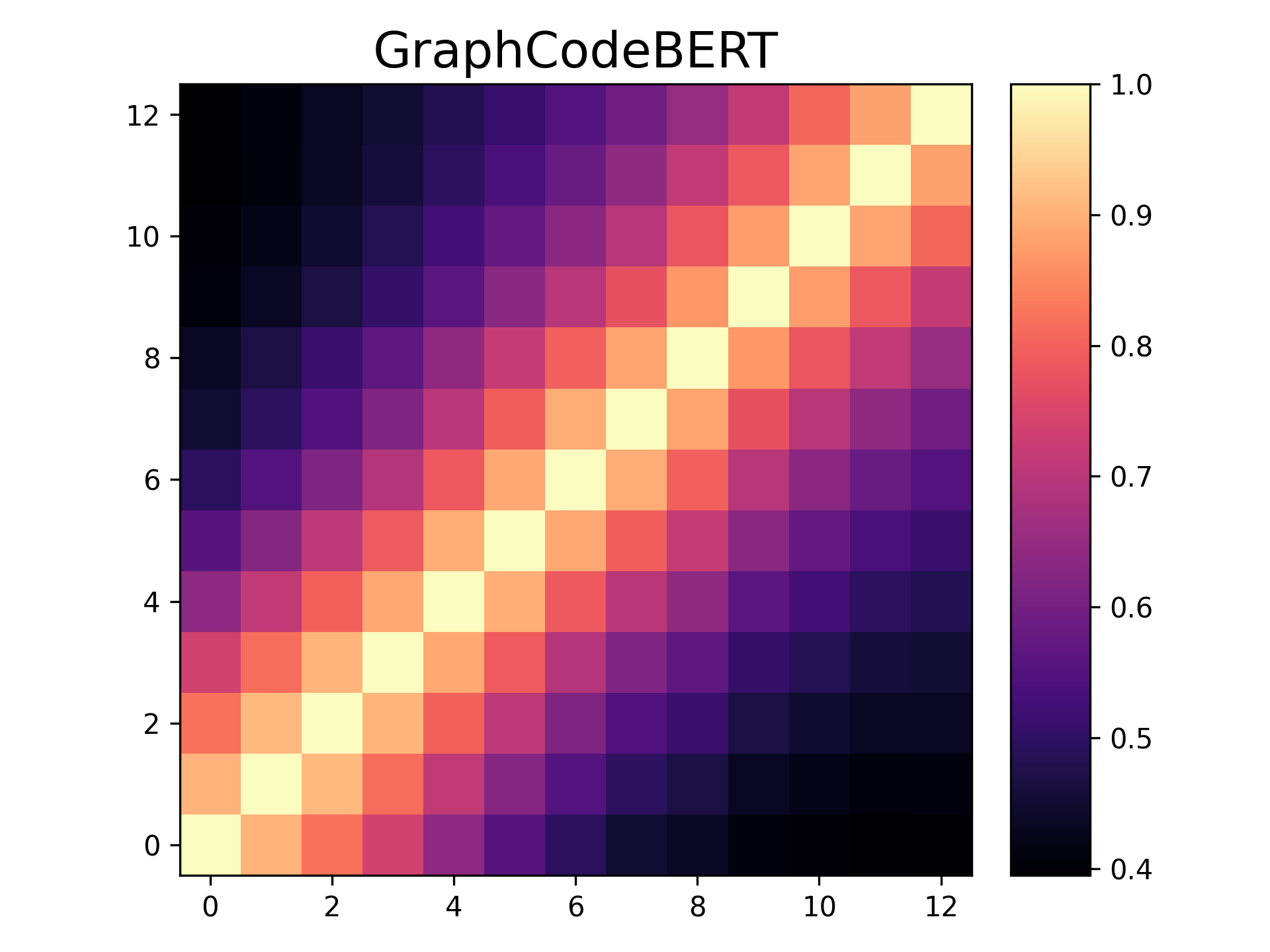}
  \label{fig:graphcodebert_cka_codesearch}
\end{subfigure}%
\begin{subfigure}{.12\textwidth}
  \centering
  \includegraphics[width=1\linewidth]{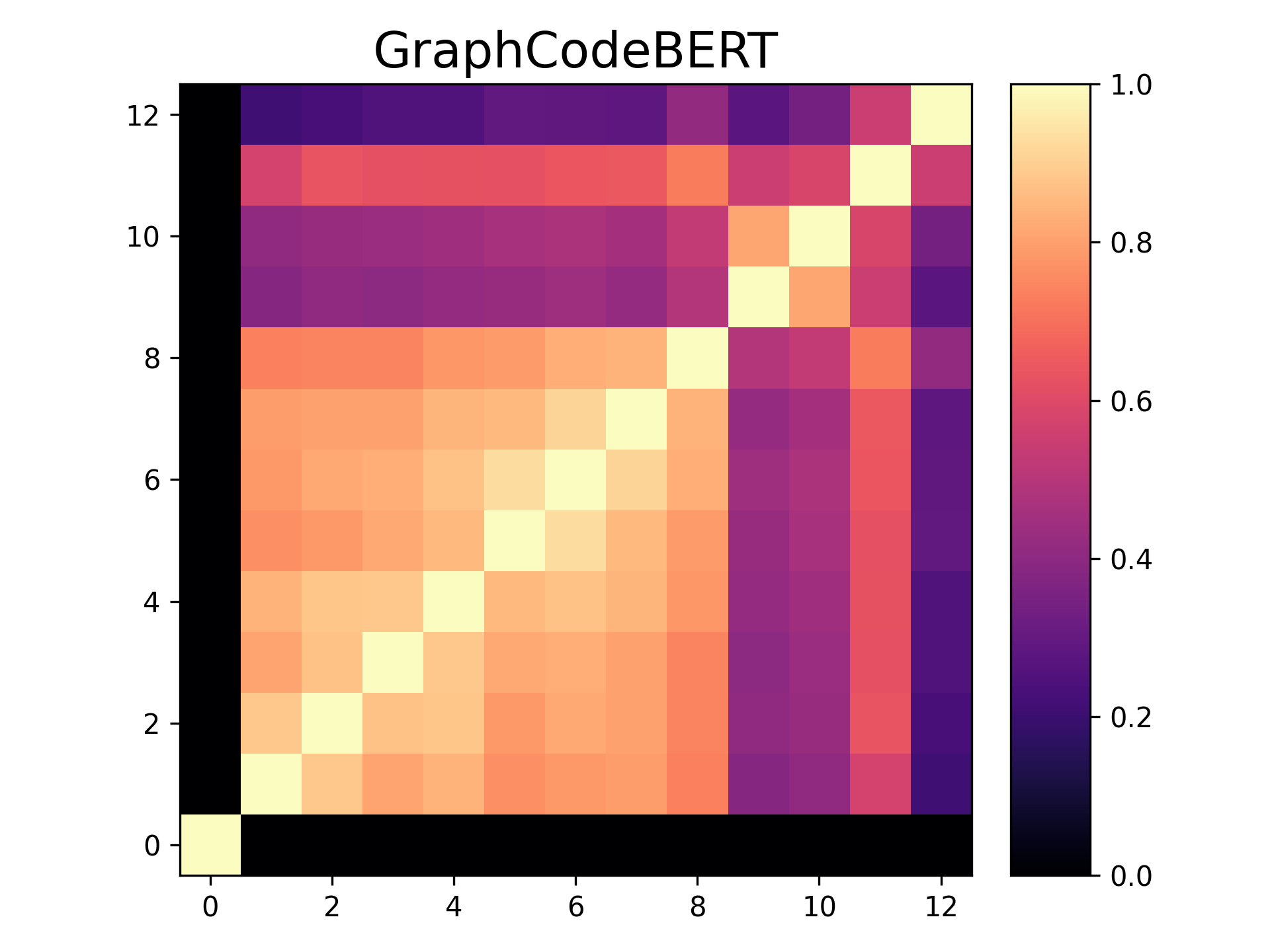}
  \label{fig:graphcodebert_cka_defdet}
\end{subfigure}%
\\
\begin{subfigure}{.12\textwidth}
  \centering
  \includegraphics[width=1\linewidth]{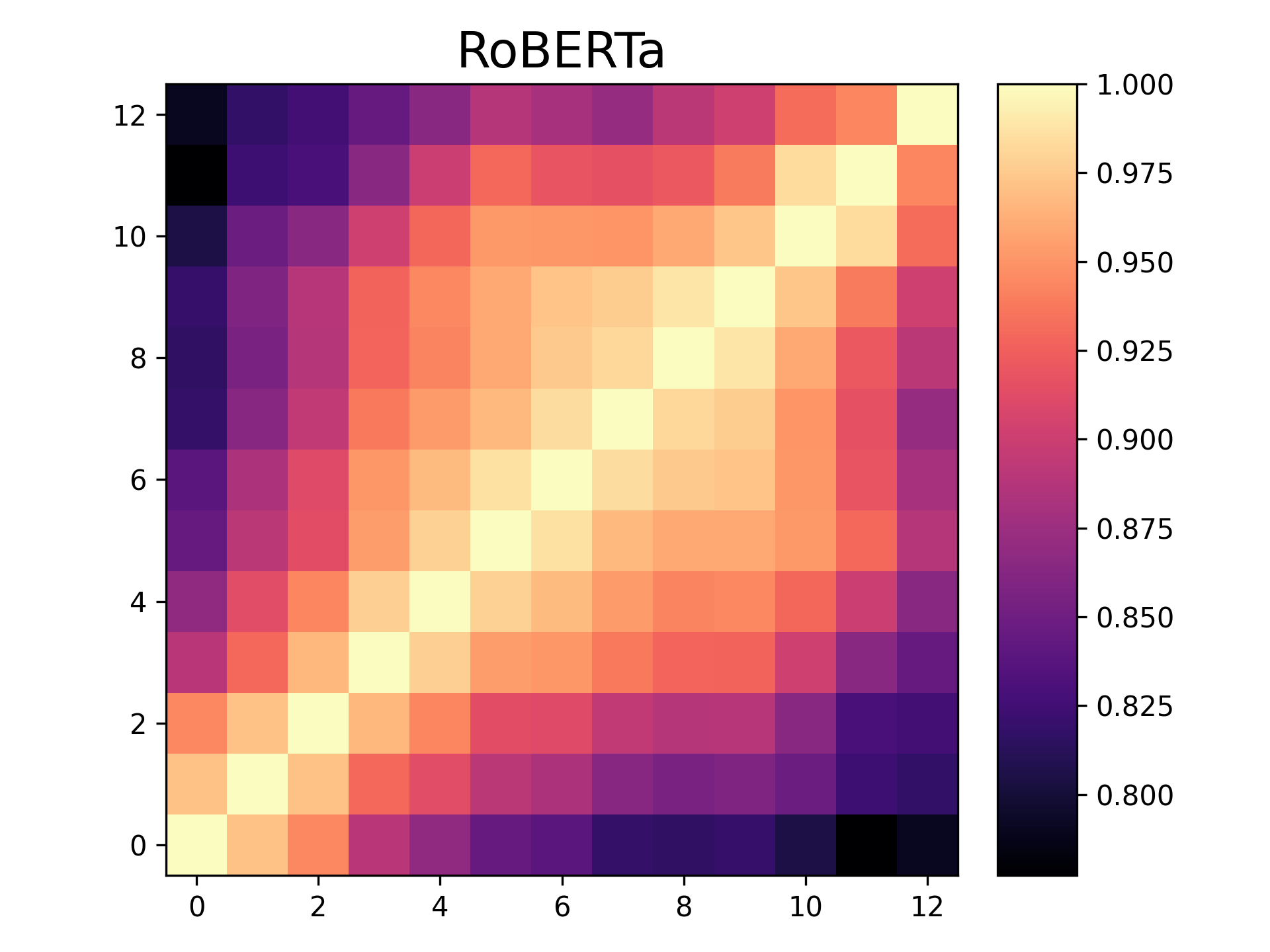}
  \label{fig:roberta_cka}
\end{subfigure}%
\begin{subfigure}{.12\textwidth}
  \centering
  \includegraphics[width=1\linewidth]{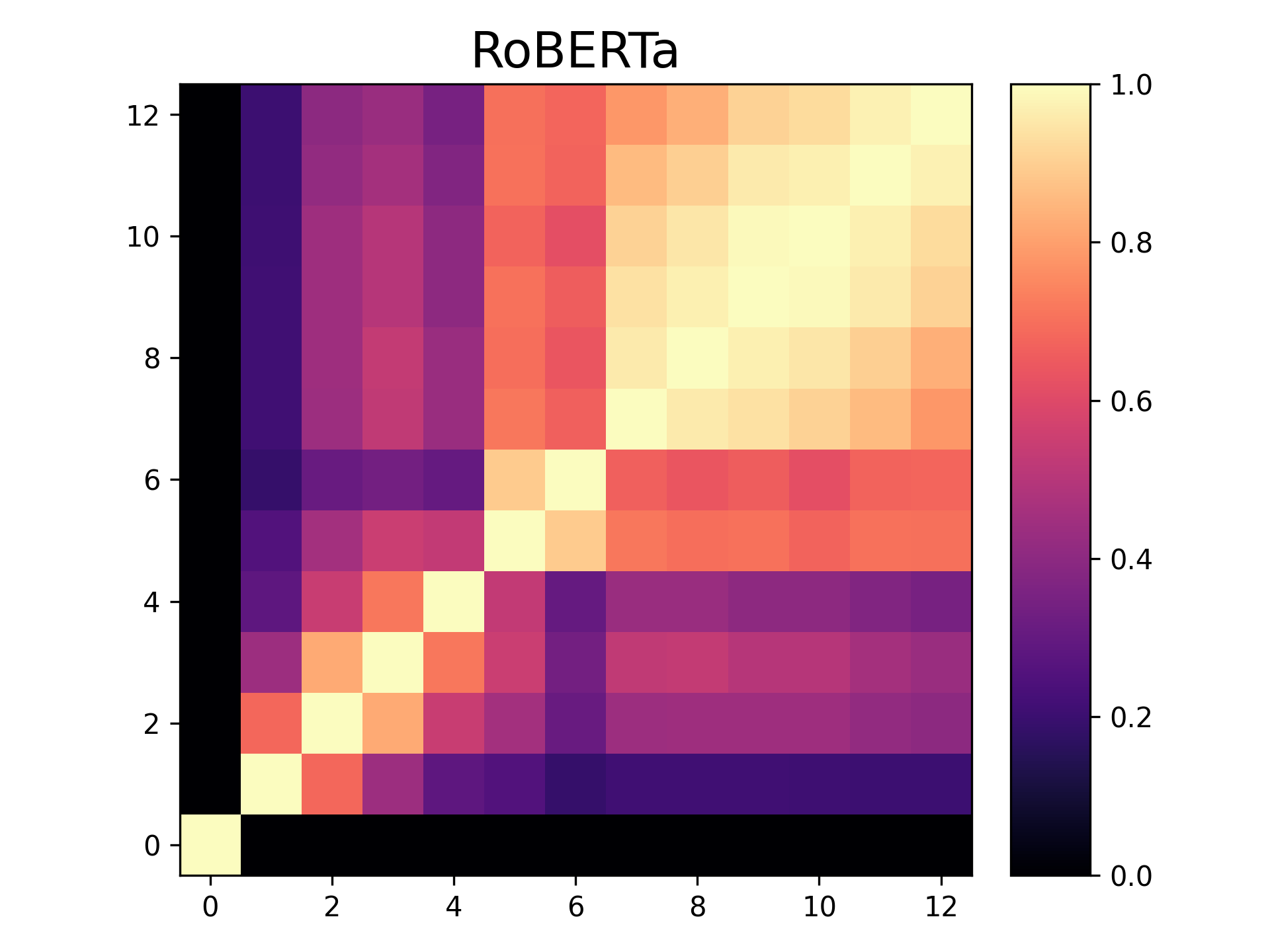}
  \label{fig:roberta_cka_clonedet}
\end{subfigure}%
\begin{subfigure}{.12\textwidth}
  \centering
  \includegraphics[width=1\linewidth]{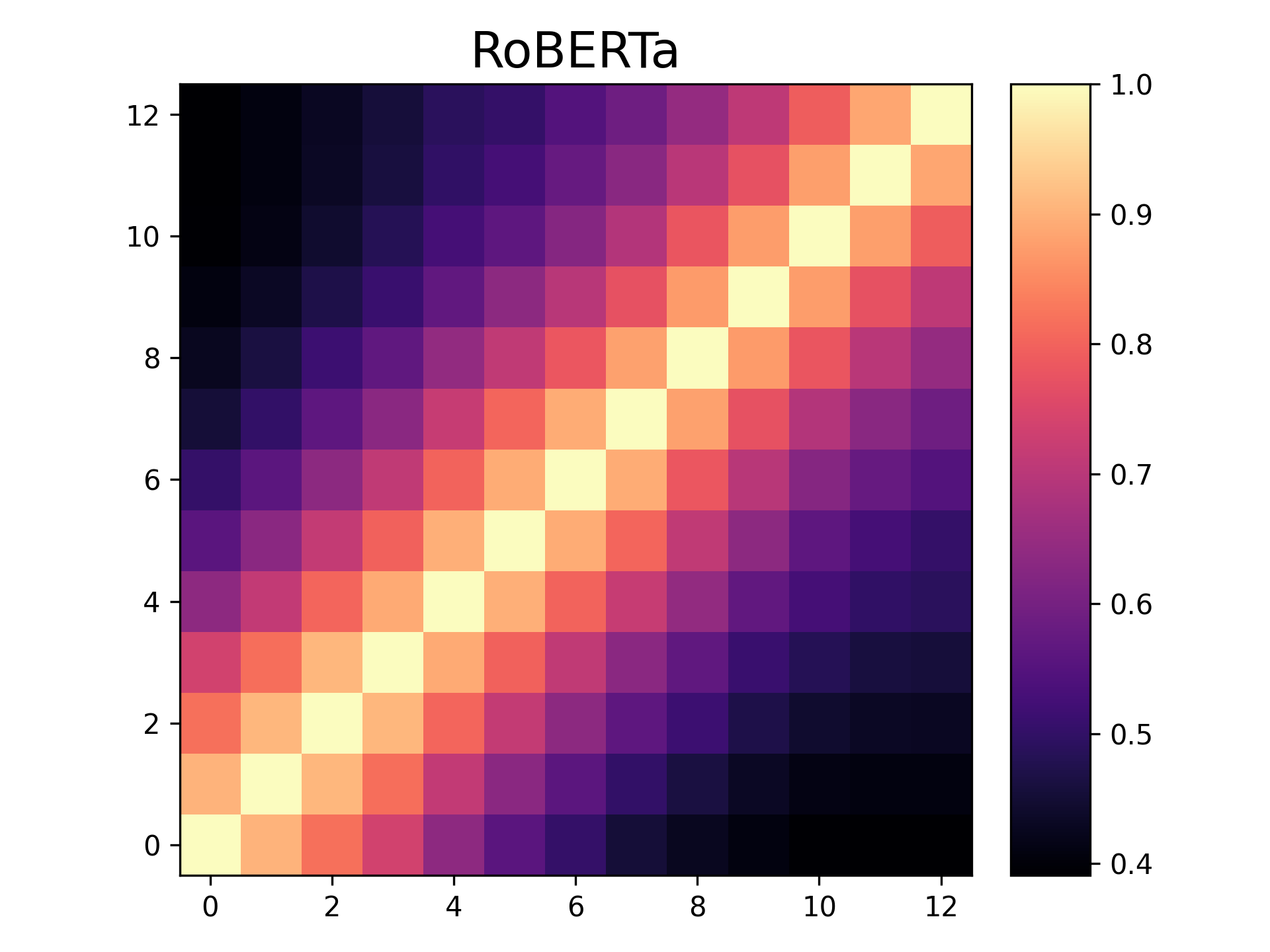}
  \label{fig:roberta_cka_codesearch}
\end{subfigure}%
\begin{subfigure}{.12\textwidth}
  \centering
  \includegraphics[width=1\linewidth]{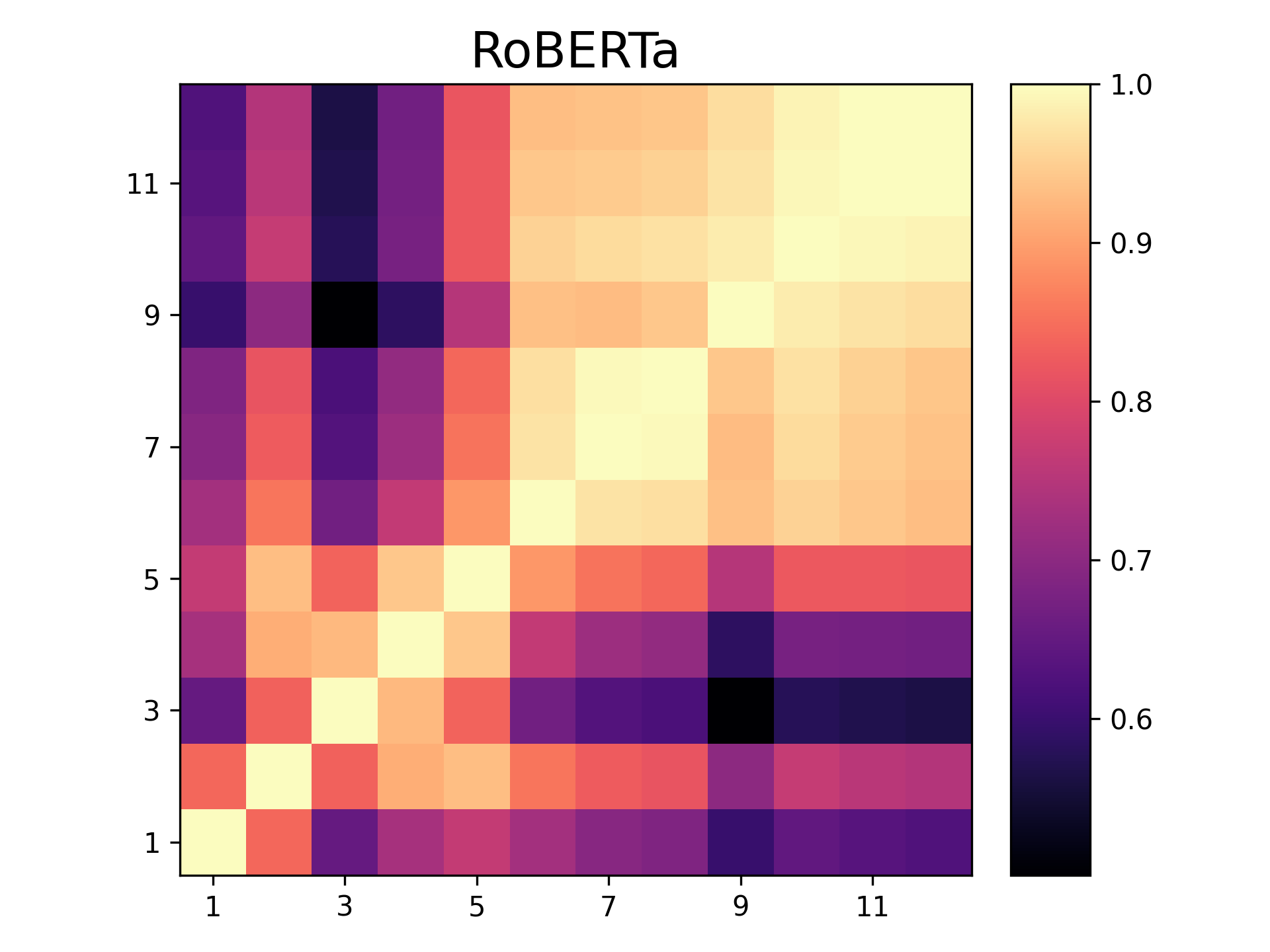}
  \label{fig:roberta_cka_defdet}
\end{subfigure}%
\\
\begin{subfigure}{.12\textwidth}
  \centering
  \includegraphics[width=1\linewidth]{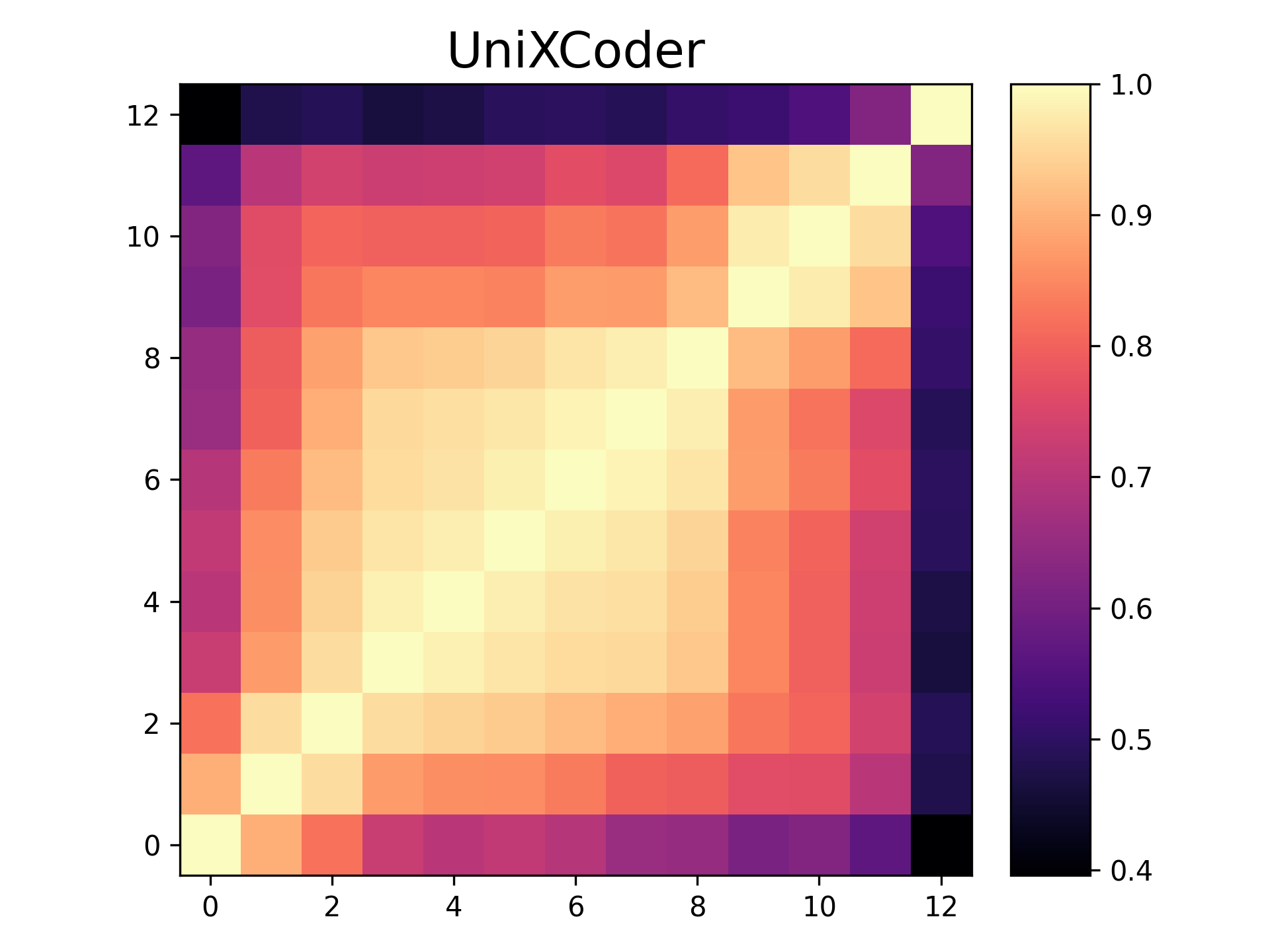}
  \caption{}
  \label{fig:unixcoder_cka}
\end{subfigure}%
\begin{subfigure}{.12\textwidth}
  \centering
  \includegraphics[width=1\linewidth]{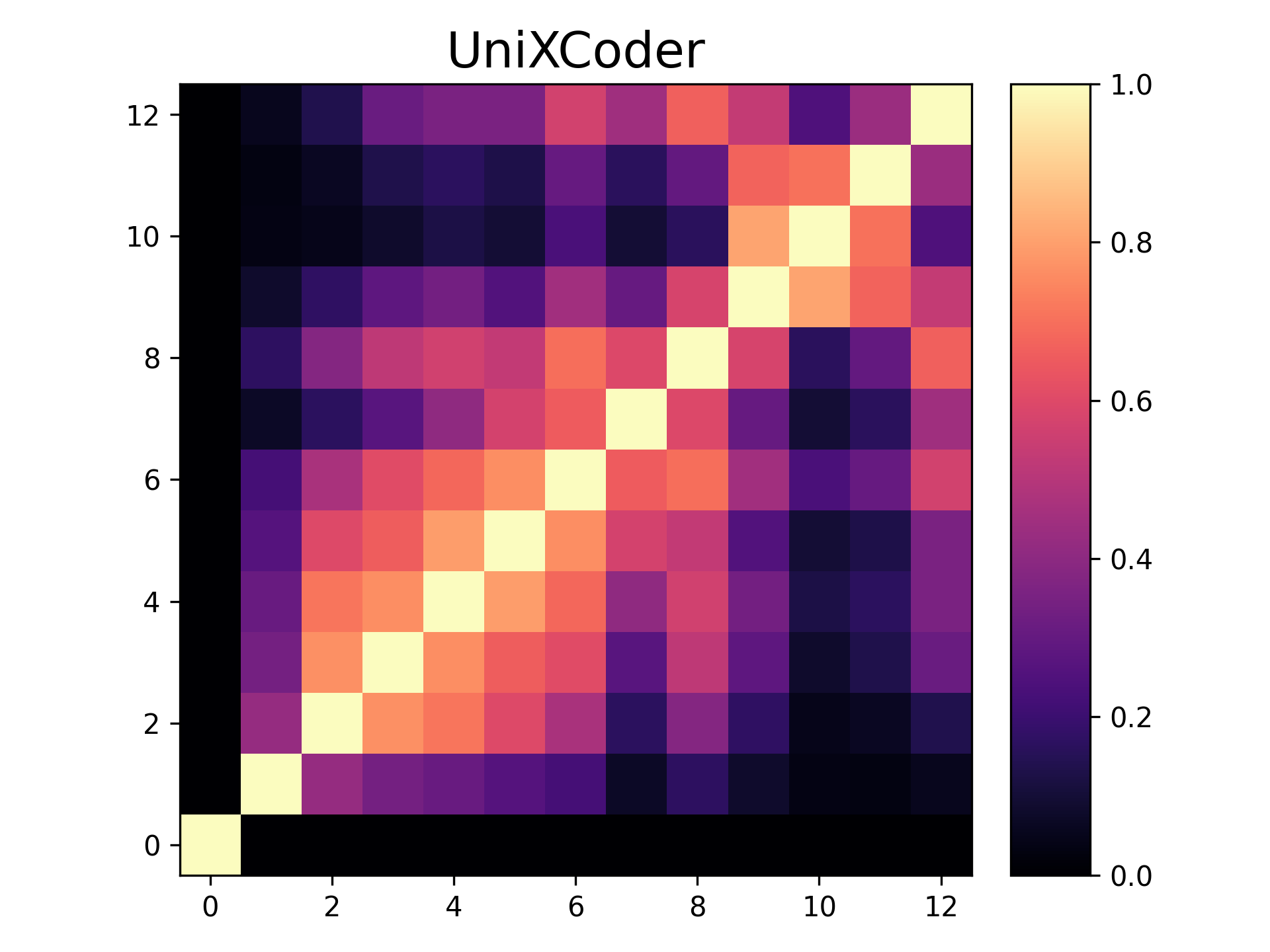}
  \caption{}
  \label{fig:unixcoder_cka_clonedet}
\end{subfigure}%
\begin{subfigure}{.12\textwidth}
  \centering
  \includegraphics[width=1\linewidth]{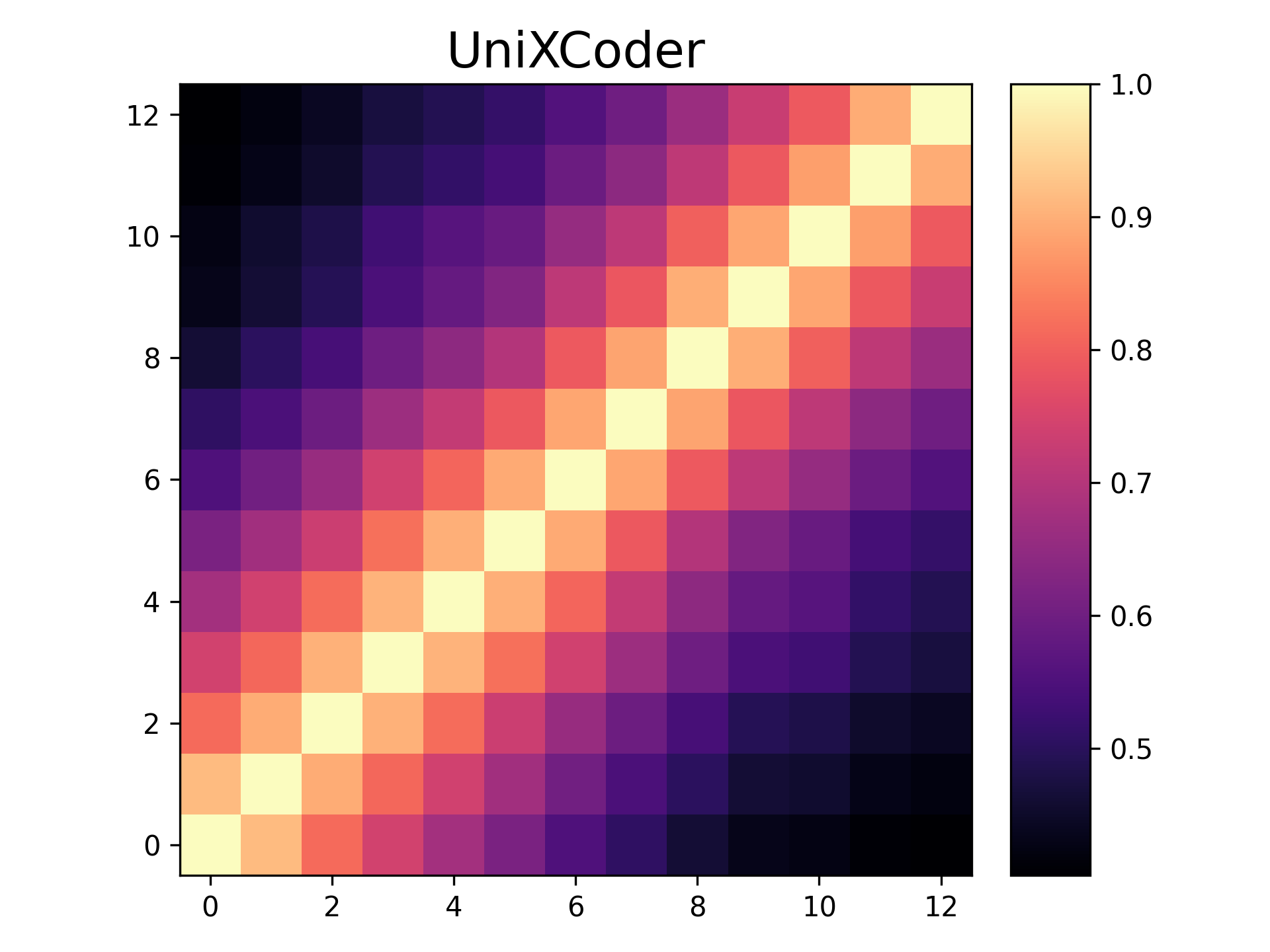}
  \caption{}
  \label{fig:unixcoder_cka_codesearch}
\end{subfigure}%
\begin{subfigure}{.12\textwidth}
  \centering
  \includegraphics[width=1\linewidth]{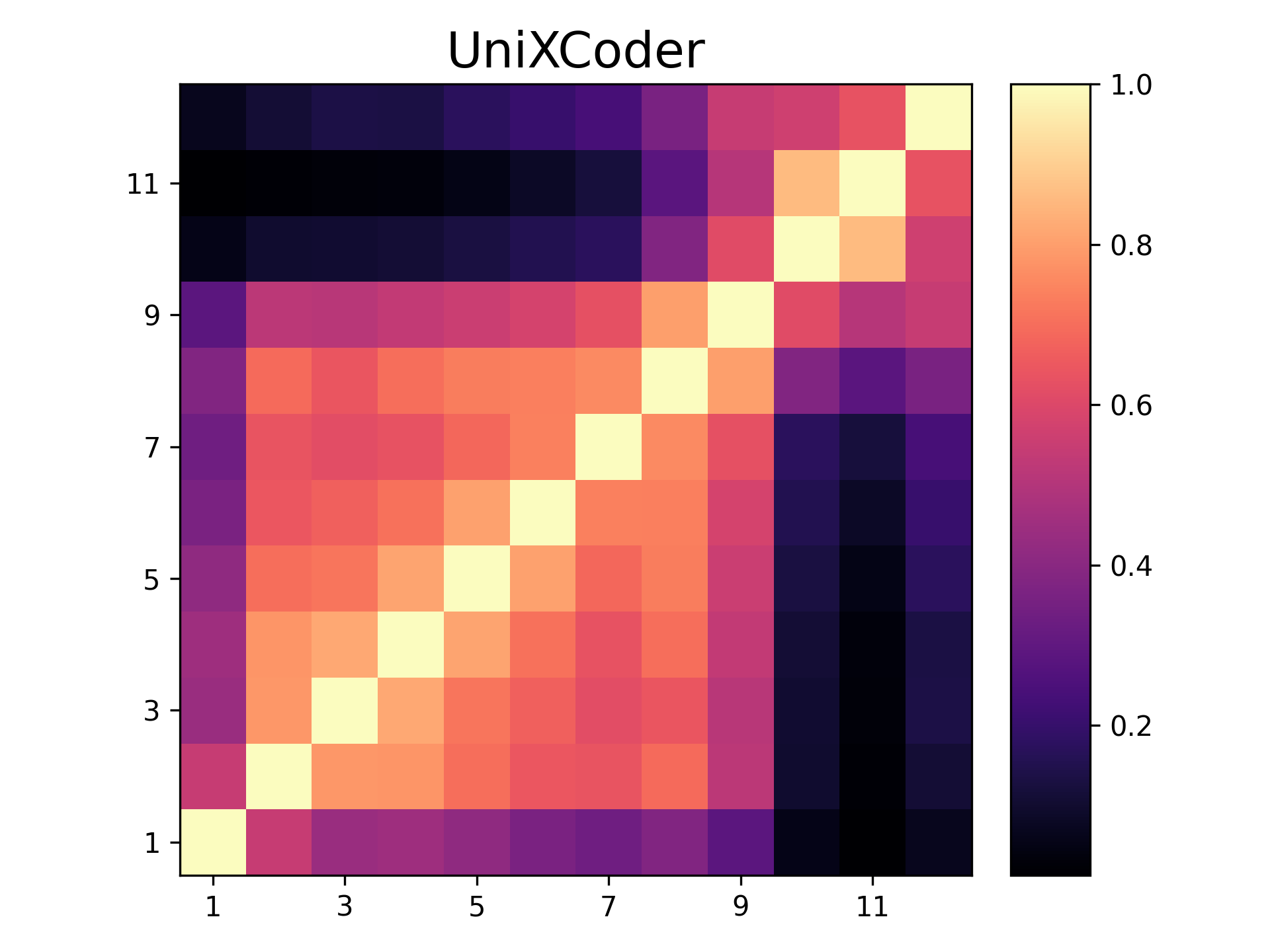}
  \caption{}
  \label{fig:unixcoder_cka_defdet}
\end{subfigure}%
\caption{Layer Redundancy Analysis, a: Token Tagging, b: Clone Detection, c: Code Search, d: Defect Detection}
\label{fig:cka_token}
\end{figure}

\subsection{Concept Analysis: Tracing specific neurons to code properties}
\label{Evaluation-Concept Analysis}
Our approach to concept learning is described in section \ref{Approach-Definition-Concept Learning}. We make the following observations. 

\begin{table}[]
\caption{Tracing important neurons to properties of code} \label{tab:top words}
\textcolor{purple}{$\blacksquare$} NUMBER \textcolor{brown}{$\blacksquare$} IDENTIFIER \textcolor{teal}{$\blacksquare$} STRING \\ \textcolor{orange}{$\blacksquare$} KEYWORD \textcolor{pink}{$\blacksquare$} MODIFIER \textcolor{green}{$\blacksquare$} TYPE\\
\begin{tabular}{|c|c|c|c}
\hline
\textbf{Neuron} & \textbf{Top-5 words}                                          & \textbf{Model}              \\
\hline
\makecell{Layer 5: 305 }  & \makecell{\textcolor{brown}{identity}, \textcolor{orange}{break}, \textcolor{brown}{interval},\\\textcolor{brown}{unit},\textcolor{orange}{class} }& CodeGPTJava  \\ 
\hline
\makecell {Layer 5:519 } & \makecell{\textcolor{purple}{79}, \textcolor{purple}{104}, \textcolor{purple}{78}, \textcolor{purple}{118}, \textcolor{purple}{99}}       & \makecell{CodeGPTPy }  \\ 
\hline
Layer 5: 246   & \makecell{\textcolor{green}{boolean}, \textcolor{brown}{CUSTOM}, \textcolor{pink}{static},\\ \textcolor{teal}{"default"}, \textcolor{teal}{"gzip"}}                                   & GraphCodeBERT      \\ 
\hline
Layer 12:654   & \textcolor{pink}{protected}, \textcolor{pink}{private}, \textcolor{purple}{64}, \textcolor{teal}{'o'}, \textcolor{teal}{'0'}                                     & RoBERTa          \\ 
\hline
Layer 0:625   & \textcolor{brown}{END}, \textcolor{green}{byte}, \textcolor{brown}{R}, \textcolor{brown}{range}, \textcolor{brown}{U}                                     & UniXCoder          \\ 
\hline
\end{tabular}
\end{table}

\subsubsection{Tracing specific neurons to java token classes}
\label{Evaluation- Concept Analysis - Tracing neurons to jave tokens}
\begin{figure}
\centering
\begin{subfigure}{.5\textwidth}
  \centering
  \includegraphics[width=1\linewidth]{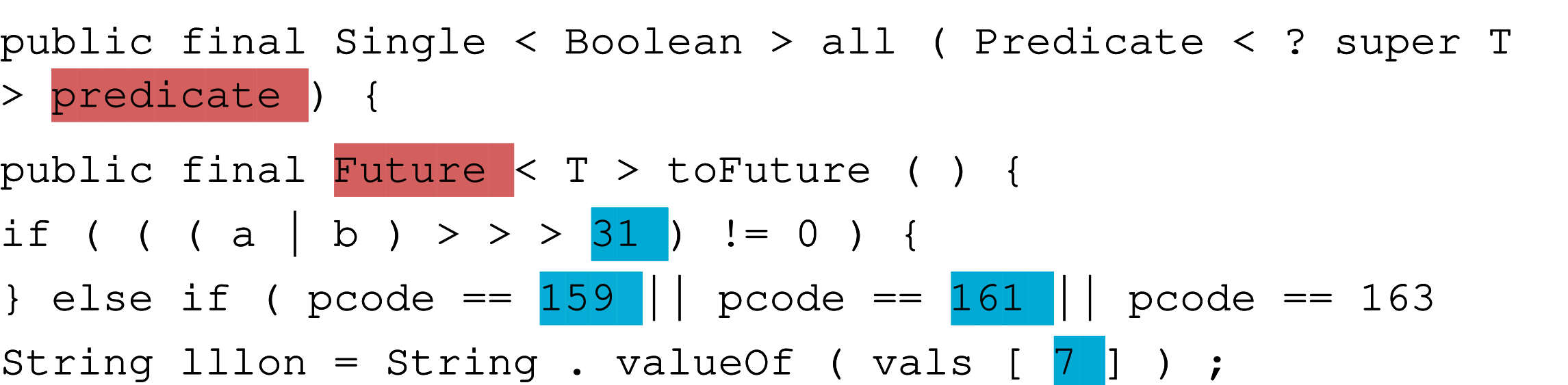}
  \caption{NUMBER, IDENTIFIER (CodeBERT) - Layer 0: 47} 
  \label{fig:activated_words_MIXTURE_NUMBER_IDENTIFIER}
\end{subfigure}%
\\

\begin{subfigure}{.5\textwidth}
  \centering
  \includegraphics[width=1\linewidth]{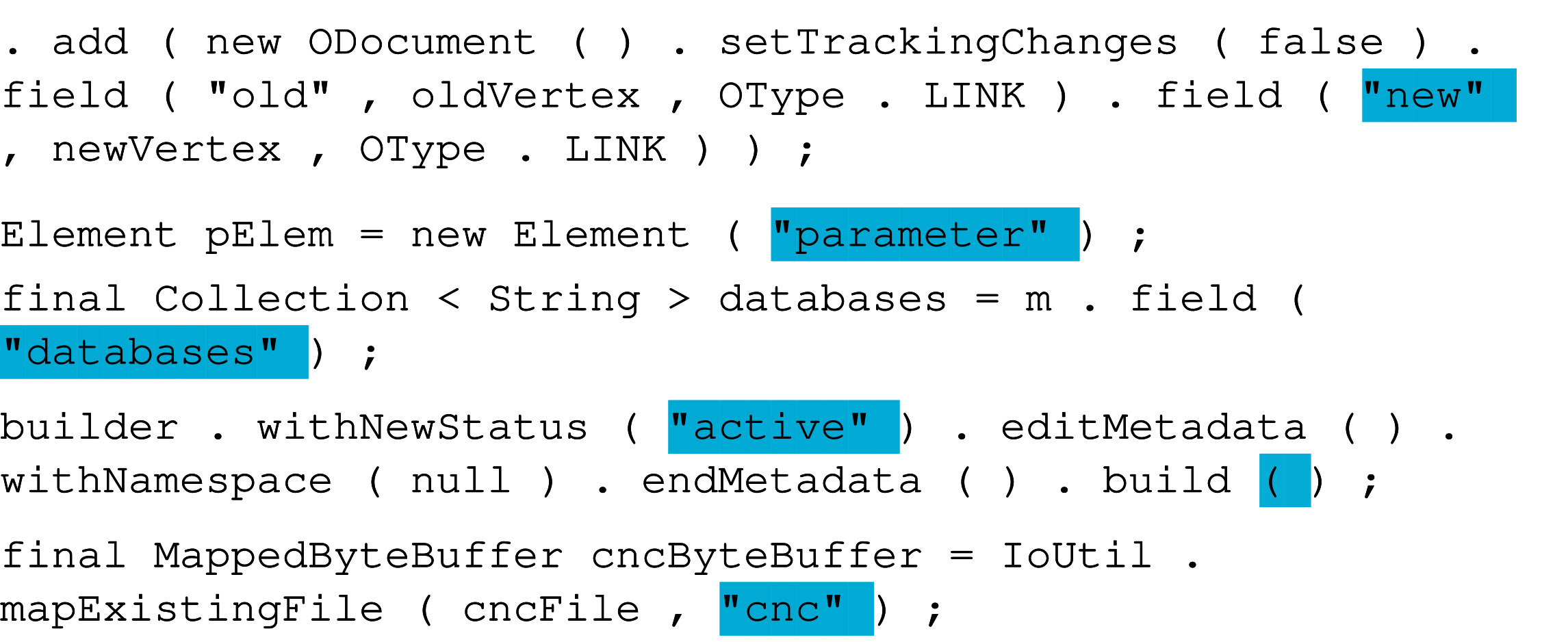}
  \caption{STRING (CodeBERT) - Layer 7: 717}
  \label{fig:activated_words_STRING}
\end{subfigure}
\caption{Activated words (Red indicates negative activation values and blue indicates positive activation values)}

\label{fig:activated_words}
\end{figure}
After reformulating our probing task based on selectivity results, we perform Linguistic Correlation Analysis (Section \ref{Background-Linguistic Correlation Analysis}) to identify the top neurons for each class. The activations of these important neurons for specific classes highlight the top words using the NeuroX \citep{dalvi2019neurox} visualization tool, as shown in Figure \ref{fig:activated_words}. Table \ref{tab:top words} presents the top words for some important neurons in five code-trained language models. We isolate clean examples of monosemanticity, polysemanticity, superposition, and composition (see: Section \ref{Approach-Definition-Concept Learning}). 

In Figure \ref{fig:activated_words_MIXTURE_NUMBER_IDENTIFIER}, neuron 47 of layer 0 in CodeBERT activates for identifiers (indicated in red with negative activation values) and numbers (indicated in blue with positive activation values). This illustrates a polysemantic neuron responsible for multiple concepts. In Figure \ref{fig:activated_words_STRING}, neuron 717 of layer 7 is primarily responsible for strings but also highlights a parenthesis. This might indicate polysemanticity or noise in the network. Further, in Table \ref{tab:top words}, Layer 5, neuron 519 for CodeGPTPython can be referred to as a 'number neuron' and is the cleanest example of monosemantic neurons we found. For CodeGPTJava, layer 5 neuron 305 is responsible for both keywords and identifiers. This could indicate superposition or be considered a monosemantic 'text neuron' that learns a higher-level concept. This observation is also supported by \citep{gurnee2023finding}. \textbf{Identifying monosemantic neurons can help us improve traceability and influence model predictions. However, superposition poses a known challenge in drawing concrete conclusions about concepts stored within neurons.} While concrete conclusions about mechanistic interpretability are hard to draw given the overlapping nature of information encoded within neurons, a deeper look into superposition and composition can help devise better interpretability studies and model compression strategies. Studies like \citep{gurnee2023finding, elhage2022toy} explore superposition, extracting features despite it and designing models with less of it, but this goes beyond the scope of our study.

In NLP, \citet{durrani2022linguistic} found clearer examples of linguistic properties associated with neurons. One limitation of probing pre-defined concepts (classes) using probing classifiers for source code tasks is the need to eliminate a significant number of deterministic tokens in our experiments (see Section \ref{Approach-Selectivity guided probing task formulation}). An approach we would like to explore in future work is focusing on model-centric concepts in an unsupervised manner. In NLP, \citet{dalvi2022discovering} identifies clusters of similar neurons within models and performs human-in-the-loop annotation to label model-centric concepts. Another approach would be to combine this analysis with extractive interpretability techniques as discussed in related work (section \ref{Related Work-Extractive}). We leave this for future work as well.

\subsubsection{Many different subsets of neurons capture important concepts}
We find that \textbf{there are multiple compositions of neurons which can predict the task with accuracy comparable to baseline.} We study some of these using layer selection and correlation clustering in the following sections, but we only analyze the best-performing ones using each method. Further analysis into the nature of superposition and different compositions of neurons storing relevant concepts can be useful for designing transfer learning, knowledge distillation, and multitask learning approaches. 



\subsubsection{Layerwise distribution of important neurons for sentence-level downstream tasks} 

Several studies about deep neural networks have suggested that lower layers learn more general features, whereas higher layers of a deep learning network store more task-specific features \citep{tenney2019bert, hao2019visualizing}. Therefore, we expected our finetuned models for sentence-level tasks to rely more on specialized features stored in the higher layers. While we observe this trend in our clone detection layerwise results (Table \ref{tab: Result2}), we find that several models can perform well by using lower layers and eliminating the higher ones without sacrificing probe accuracy. For example, in defect detection (Table \ref{tab: Results1}), RoBERTa (0-5 layers) can rely on lower layers and actually experiences an increase in accuracy (Table \ref{tab: Results1}) through layer selection. Similarly, in code search (Table \ref{tab: Result2}),  GraphCodeBERT uses 0-1 layers, and CodeGPT-Py uses 0-2 layers to achieve baseline probing accuracy or higher. This suggests that \textbf{some models can learn higher sentence-level concepts  in the initial few layers and effectively transmit them to the remaining layers.} It could also imply that they rely on more general features, usually stored in lower layers, to make predictions about these tasks. These models make good candidates for model compression.

\paragraph{Distribution of task-relevant information in neurons with similar activation patterns (based on Correlation Clustering)}

In Section \ref{Background-Correlation Clustering}, which describes correlation clustering, we mention that a high value of the clustering hyperparameter $c$ indicates larger clusters, and neurons that are further apart (less similar to each other) are included in the independent set, which is then eliminated. For most models used in defect detection and code search tasks, a value of $c$ above 0.5 has been selected, while several models in token tagging and clone detection tasks have lower selected thresholds.

Models with a lower clustering threshold (highlighted in yellow) result in the elimination of fewer neurons due to their similarity, as these neurons contain task-relevant information. This suggests that task-related information is distributed across neurons that demonstrate very similar activation patterns. Despite containing overlapping information, these neurons hold pertinent details that contribute to decreased accuracy when they are eliminated, as observed in token tagging and clone detection tasks. On the contrary, in defect detection and code search models, a large number of similar neurons can be eliminated as they do not contain task-relevant information used in the probing experiments. Generally, a low clustering threshold leads to a lower percentage reduction in neurons. However, it is interesting to note that the CodeGPT-Java and CodeGPT-Python models for clone detection show a high percentage reduction in neurons despite a low clustering threshold. This suggests that a significant number of neurons have very similar activation patterns and can be eliminated. This observation aligns with other results (see: CKA results for CodeGPT-Java and CodeGPT-Python in Figure \ref{fig:cka_token}), indicating high similarity within these models, likely due to their training on monolingual data. \textbf{This suggests that high-similarity by itself does not indicate that layers can be pruned or are not important, and it is worthwhile to analyze how task-related concepts are stored to select relevant parts of the network while devising strategies for creating more efficient and interpretable models.}

\subsubsection{Implications for interpretability of latent representations and input-neuron traceability}
\label{Evaluation- Implications-traceability}
 Our experiments also suggest that a very small set of neurons can predict the given task with an accuracy comparable to oracle accuracy. Isolating important subsets of neurons also leads to better traceability, allowing us to modify neurons consistently correlated with incorrect predictions. However, whether the task truly requires very few neurons to make predictions or if it relies on shortcuts as suggested by \citep{rabin2022syntax} requires further investigation. If the model does rely on shortcuts, our work motivates the following questions: Did the model not learn anything about the important features? Did the latent representations encode relevant features during self-supervised pretraining that are not being used to make predictions for the specific task? If so, are there better code-specific training objectives, as suggested by \citet{troshin2022probing}'s probing study, or inference time neuron modifications that can mitigate this?

Further experiments on tasks of different granularities can also help us identify patterns within subsets of neurons responsible for particular properties. Our neuron analysis can serve as a link between \textit{extractive} interpretability approaches (discussed in \cref{Related Work-Extractive}) and studies of latent code representations (\cref{Related Work-embeddings}). It can help trace relationships between input features and important neurons, enabling interpretation and control of predictions. In NLP studies, neurons can be modified to control biases like gender, race, etc. Similar approaches could be especially useful for use cases like vulnerability detection, where neuron modification might correct incorrect vulnerability signals in the model. This is a promising research direction that we leave for future work.

\section{Threats to Validity}
\label{Threats to Validity}

We ensure software quality by using the models and datasets shared by the original developers under similar experimental settings as much as possible. Each of the models and datasets used has been used and cited by several studies in machine learning for software engineering. However, differences due to pretraining settings or decisions made by them cannot be accounted for while comparing the results of different models. Our results may not generalize to other code-related tasks. Further, while probing classifiers are a well-established interpretability method, the performance of probing classifiers accuracy is contingent on several possible confounders like the original data set, pre-trained model, probing data set, and probing model, which make it hard to make absolute claims about representation quality \citep{belinkov2022probing}.  Another potential threat is that  neuron ranking methods may not always be optimal. \citet{antverg2021pitfalls}  point out that a good probing classifier may perform well even when the ranking is bad, or an average classifier on a good ranking may perform better than a good classifier on a bad ranking. Therefore the probe-ranking combination requires further investigation and possible adaptation for code-related models and tasks. 

\section{Conclusion and Future Work}
\label{Conclusion and Future work}

In this work, we conduct a comprehensive neuron-level analysis of seven code-trained language models across four software engineering tasks. To our knowledge, no other works have analyzed SE downstream tasks and models in such a fine-grained manner.  Our approach leverages probing classifiers, and we propose using selectivity to guide the formulation of probing tasks, considering memorization due to degeneracies in source code datasets. We use neuron similarity and ranking techniques and perform a redundancy analysis by eliminating redundant neurons due to similarity and irrelevance to the task, followed by concept analysis to examine the concentration and distribution of important neurons within the latent representations. This helps us understand how transferable and decomposable concepts within latent representations are. 

We discuss concrete trends for specific models and tasks and provide design guidelines if there are no overarching trends for devising strategies for multitask learning, knowledge distillation, and model compression and improving traceability between inputs, neurons, and outputs to develop more efficient and interpretable code-trained models.

We also provide directions for future research for others, like exploring superposition in neurons, as that is a hard problem in itself and requires a much more detailed set of case studies and experiments as evidenced by \citet{gurnee2023finding} and is beyond the scope of our neuron analysis. Our work also lays the groundwork for future work related to mechanistic interpretability of code-trained models (section \ref{Evaluation- Concept Analysis - Tracing neurons to jave tokens}), identifying spurious correlations and influencing model predictions (section \ref{Evaluation- Implications-traceability}), model decomposition and pruning (section \ref{Evaluation- Implicaitons model distillation and pruning}).

\bibliographystyle{ACM-Reference-Format}
\bibliography{references}

\appendix

\end{document}